\documentclass[reqno,12pt]{article}
\usepackage{amsfonts,amssymb,amsmath,amscd,bm,bbold,cite,delarray,subfigure}
\usepackage{xypic}
\usepackage[dvips]{color}
\usepackage[dvips]{graphicx}

\numberwithin{equation}{section}

\DeclareMathOperator{\gh}{gh}

\font\capital=rsfs12
\font\scriptcapital=rsfs10 at 7 truept
\font\scriptscriptcapital=rsfs10 at 5 truept
\font\sansserif=cmss12
\font\scriptsansserif=cmss12 at 7 truept
\font\scriptscriptsansserif=cmss10 at 5 truept
\font\euler=eusm10 at 12 truept
\font\scripteuler=eusm7
\font\scriptscripteuler=eusm5 
\newcommand{\bfs}[1]{{\boldsymbol{#1}}}

\begin{document}

\hrule\vskip.4cm
\hbox to 14.5 truecm{January 2008\hfil DFUB 08}
\hbox to 14.5 truecm{Version 1  \hfil } 
\vskip.4cm\hrule
\vskip.7cm
\begin{large}
\centerline{\textcolor{blue}{\bf Gauging the Poisson sigma model} }  
\centerline{\textcolor{blue}{\bf }}  
\end{large}
\vskip.2cm
\centerline{by}
\vskip.2cm
\centerline{\textcolor{blue}{\bf\bf Roberto Zucchini}}
\centerline{\it Dipartimento di Fisica, Universit\`a degli Studi di Bologna}
\centerline{\it V. Irnerio 46, I-40126 Bologna, Italy}
\centerline{\it I.N.F.N., sezione di Bologna, Italy}
\centerline{\it E--mail: zucchinir@bo.infn.it}
\vskip.7cm
\hrule
\vskip.7cm
\centerline{\textcolor{blue}{  \bf Abstract} }
\par\noindent
\vskip.4cm
We show how to carry out the gauging of the Poisson sigma model in an 
AKSZ inspired formulation by coupling it to a generalization of the 
Weil model worked out in ref.
\cite{Zucchini6}. We call the resulting gauged field theory, Poisson--Weil
sigma model. We study the BV cohomology of the model and show its relation to
Hamiltonian basic and equivariant Poisson cohomology. 
As an application, we carry out the gauge fixing of the pure 
Weil model and of the Poisson--Weil model. In the first case, we obtain the
$2$--dimensional version of Donaldson--Witten topological gauge theory, 
describing the moduli space of flat connections on a closed surface.
In the second case, we recover the gauged $A$ topological sigma model worked out
by Baptista describing the moduli space of solutions of the so--called 
vortex equations.

\par\noindent
Keywords: quantum field theory in curved spacetime; geometry, 
differential geometry and topology.
PACS: 04.62.+v  02.40.-k 

\vfill\eject

\begin{small}
\section{\bf Introduction} 
\label{sec:intro}
\end{small}

One efficient way of generating a sigma model on a non trivial manifold $X$
is the gauging of a sigma model on a simpler manifolds $M$ carrying the action of
a Lie group $G$ such that $X\simeq M/G$. The target space 
of the gauged model turns out to be precisely $X$. 
In many interesting cases, a symplectic structure on $M$ and a moment map for the 
$G$--action can be defined and this construction is a particular
case of a general procedure called Hamiltonian reduction \cite{Marsden:1}.

The usefulness of gauging sigma models 
was first recognized by Witten in \cite{Witten3}, where the gauged linear sigma-model
with target $X=\mathbb{C}^n$ and group $G=U(1)$ was used to study non-gauged 
sigma-models into weighted projective spaces and Calabi-Yau hypersurfaces thereof. 
Later, in \cite{Witten4}, applying the same procedure, Witten employed 
a gauged linear sigma-model with target $X=\mathbb{C}^{kn}$ and group $G=U(k)$ 
in a study of the quantum cohomology of Grassmannians. 

The study of gauged sigma models, however, was initiated long before
Witten's work. Developing on the results of Gates, Hull and Ro\v cek 
in \cite{Gates1}, the gauging of (2,2) supersymmetric sigma models on 
biHermitian manifolds was studied originally by Hull, Papadopoulos and 
Spence in \cite{Spence1}.
Their analysis was however limited to the subclass of almost product
structure target spaces because of the lack of an off--shell 
(2,2) supersymmetric action in the general case at that time. 
After the realization that biHermitian geometry is naturally framed in 
generalized complex and Kaehler geometry by Hitchin and Gualtieri
\cite{Hitchin1,Gualtieri}, (2,2) supersymmetric sigma models have been
fruitfully formulated in this new powerful geometric language.
In this way, the off--shell (2,2) supersymmetric sigma model action
on general biHermitian manifolds was recently obtained in ref. \cite{Lindstrom5}.
This has led the authors of ref. \cite{Zayas1} to extend the analysis
of \cite{Spence1} to general biHermitian target spaces. 
In \cite{Kapustin3}, the same analysis has been carried out in 
the on--shell formalism. 

(2,2) supersymmetric sigma models are rather complicated quantum field theories
and, so, they are difficult to study. In 1988, Witten showed that a (2,2)
supersymmetric sigma model on a Calabi--Yau manifold (a particular case of
biHermitian manifold) could be twisted in two different ways,   
to give the so called $A$ and $B$ topological sigma models \cite{Witten1,Witten2}.  
Unlike the original untwisted sigma model, the topological models are soluble:
the calculation of observables can be reduced to classical problems of geometry.
Topological sigma models on general biHermitian manifolds have been
worked out in recent years to a various degree of depth in
\cite{Kapustin1,Kapustin2,Zucchini4,Zucchini5,Chuang}. 
However, only a small number of papers has been devoted to the study of 
gauged topological sigma models \cite{Baptista1,Baptista2,Baptista3} and these
are concerned with the Calabi--Yau case only. The problem arises of
constructing gauged topological sigma models with more general biHermitian target space
geometries. 

In the last few years, many attempts have been made to construct topological
sigma models with generalized complex and Kaehler target manifolds 
\cite{Lindstrom2,Lindstrom3,Zucchini1,Zucchini2,Pestun1,Guttenberg}.
In \cite{Zucchini1,Zucchini2,Pestun1}, the sigma models were worked out by 
employing the Batalin--Vilkovisky (BV) quantization algorithm \cite{BV1,BV2} in  
the Alexandrov--Kontsevich--Schwartz--Zaboronsky (AKSZ) formulation \cite{AKSZ}. 
To date, this seems to be the most promising approach to the solution of 
the problem, though, as shown in \cite{Zucchini3}, the implementation 
of gauge fixing remains a major technical obstacle even in the simplest cases.  

In ref. \cite{Zucchini6},  we showed how Hamiltonian symmetry reduction 
could be incorporated in the sigma model on generalized complex manifolds 
worked out in refs. \cite{Zucchini1,Zucchini2} (the so--called Hitchin model). 
This was achieved by coupling the sigma model 
to a kind of ghostly Poisson sigma model called Weil model.
To illustrate our procedure, we applied it also to the 
standard Poisson sigma model \cite{Ikeda2,Strobl}
in the AKSZ formulation of refs. \cite{Cattaneo1, Cattaneo2}.

As it turns out, coupling to the Weil model amounts to a gauging procedure.
In \cite{Zucchini6}, the background principal bundle was taken to be trivial.
In this paper we show that this restriction is not in any way essential. 
With appropriate modifications, the Weil sigma model can be formulated 
and the coupling of the Weil model to the relevant 
sigma model can be implemented for a general principal bundle. 
We restrict ourselves to the Poisson sigma model for its simplicity and 
its independent interest. 
Our construction results in a gauged Poisson sigma model, 
which we call Poisson--Weil sigma model. 

It is instructive to write down the classical action of the Poisson--Weil sigma model
to see its relation to the
conventional formulation of standard Poisson sigma model. The target space is a Poisson 
manifold $M$ with Poisson structure $P^{ab}$ carrying a Hamiltonian action of 
a Lie group $G$ with fundamental vector field $u_i$ and moment map $\mu_i$ and
leaving $P^{ab}$ invariant. The base space $\Sigma$ supports a principal 
$G$--bundle $Q$. The fields are an embedding field $x^a$, a 
cotangent space valued $1$--form field 
$\eta_a$, as in the ordinary Poisson model, and a gauge field $A^i$,
a coadjoint scalar field $b_i$ and an adjoint scalar field $B^{+i}$. 
The classical action is
\begin{equation}
S=\int_\Sigma\Big[
-b_iB^{+i}-b_iF_A{}^i-\mu_i(x)B^{+i}
+\eta_aD_Ax^a+\hbox{$\frac{1}{2}$}P^{ab}(x)\eta_a\eta_b\Big],
\end{equation}
where \hphantom{xxxxxxxxxxxxxxxxxxxxxxxxxxxxxxx}
\begin{align}
&F_A{}^i=dA^i+\hbox{$\frac{1}{2}$}f^i{}_{jk}A^jA^k,
\\
&D_Ax^a=dx^a-u_i{}^a(x)A^i
\end{align}
are the gauge curvature of $A^i$ and the gauge covariant derivative of 
$x^a$, respectively.
The Poisson--Weil sigma model 
enjoys a large 
symmetry which extends that of the ordinary 
Poisson sigma model by the gauge symmetry. The symmetry closes only on shell,
as in the ordinary case. This disease is cured by using a suitable BV
formulation generalizing that of \cite{Cattaneo1, Cattaneo2}.

The Weil and Poisson--Weil sigma models have a very rich algebraic and
geometric structure. The BV cohomology of the Weil model is related to the  
basic cohomology of the Weil algebra $W(\mathfrak{g})$ of the Lie algebra
$\mathfrak{g}$ of $G$ (in turn isomorphic to the de Rham cohomology  
of the classifying space $BG$ of the group $G$). The BV cohomology of the
Poisson--Weil model is related to the Hamiltonian basic and equivariant Poisson cohomology
of the Poisson manifold. To some extent, this is expected on general grounds 
and the fact that this is indeed so shows the soundness of the models. 

As an application, we work out the gauge fixing of the pure Weil sigma model 
and of the Poisson--Weil sigma model in the BV framework. In the first case, we obtain the
$2$--dimensional version of Donaldson--Witten theory, a topological field theory  
describing the moduli space of flat connections 
on a closed surface \cite{Witten5,Witten6}. In the second case, we recover the gauged
topological sigma model worked out by Baptista in refs.
\cite{Baptista1,Baptista2,Baptista3}, which describes the moduli space of
solutions of the so--called vortex equations and is a gauged version of
Witten's $A$--model \cite{Witten1,Witten2}.

The plan of the paper is as follows.
In sect. \ref{sec:yangmills}, we present a generalization of the Weil sigma model
originally worked out in ref. \cite{Zucchini6}, which is valid for a general
principal $G$--bundle on the sigma model world sheet and is suitable
for the constructions of the following sections. 
In sect. \ref{sec:yangmillsfixing}, we work out a gauge fixing of the Weil
model and show that it yields the $2$--dimensional version of Donaldson--Witten theory.
In sect. \ref{sec:gaugedpoisson}, we formulate a generalization of the
Poisson--Weil sigma model worked out in ref. \cite{Zucchini6} and show that it
constitutes a gauging of the ordinary Poisson model.
In sect. \ref{sec:poissonweilfixing}, we carry out the gauge fixing of the
Poisson--Weil sigma model and show that it reproduces the gauged
topological sigma model by Baptista. In sect. \ref{sec:outlook},
we outline briefly potential generalizations of the 
constructions of this paper to the case where $G$ is a Poisson--Lie group. 
Finally, in the appendices, we conveniently collect 
various relations and identities which may help the reader
willing to check the details of our analysis.

\vskip 1cm

{\bf Acknowledgments.} We thank F. Bonechi, T. Strobl, H. Bursztyn 
for helpful discussions. We thank also, J.-P. Ortega, R. L. Fernandes,
V. Ginzburg and E. Meinrenken for correspondence. 

\vskip 1cm

\vfill\eject

\begin{small}
\section{  \bf The Weil sigma model}
\label{sec:yangmills}
\end{small}

In this section, we present a generalization of the Weil sigma model
originally worked out in ref. \cite{Zucchini6}, which is suitable
for our construction.  
Though the covariance of the superfields of the version of Weil sigma model
studied here is more general, its BV formulation is essentially the same as
that of \cite{Zucchini6}. The reader is therefore invited to read that paper 
for more details on the BV formalism used and the derivation of the classical 
master equation and BV variations below. 

We consider a geometrical setting consisting of the following elements.

\vskip.27cm

\begin{enumerate}

\item A closed surface $\Sigma$.

\item A compact connected Lie group $G$ with Lie algebra $\mathfrak{g}$.

\item A principal $G$--bundle $Q$ over $\Sigma$.

\end{enumerate}

With $\Sigma$ there is associated the degree shifted tangent bundle
$T[1]\Sigma$. Let $a_1:T[1]\Sigma\rightarrow \Sigma$ be the associated bundle 
projection. Then, we can construct the pull-back principal bundle 
$a_1{}^*Q$ over $T[1]\Sigma$. Concretely, $a_1{}^*Q$ can be described in the
language of $1$ cocycles as follows.
Let $\{U_A\}$ be an open covering of $\Sigma$
such that $Q|_{U_A}\simeq U_A\times G$. Let $\{g_{AB}\}$ be the 
$G$--valued $1$--cocycle representing $Q$ with respect to the covering
$\{U_A\}$. Define $\bfs{g}_{AB}=g_{AB}\circ a_1$. Then, $\{\bfs{g}_{AB}\}$ 
is the $G$--valued $1$--cocycle representing $a_1{}^*Q$ with respect to 
the covering $\{a_1{}^{-1}(U_A)\}$ of $T[1]\Sigma$. 

A generalized connection $\bfs{c}$ of $a_1{}^*Q$  
is defined as follows. $\bfs{c}$ is given 
locally on each open set $a_1{}^{-1}(U_A)$ of $T[1]\Sigma$
as a function $\bfs{c}_A\in\Gamma(a_1{}^{-1}(U_A),\mathfrak{g}[1])$ with
$\bfs{c}_A=\mathrm{Ad}\,\bfs{g}_{AB}\,\bfs{c}_B-\bfs{g}_{AB}\bfs{d}(\bfs{g}_{AB}{}^{-1})$
on $a_1{}^{-1}(U_A)\cap a_1{}^{-1}(U_B)\not=\emptyset$, where 
$\bfs{d}$ is the homological vector field of $T[1]\Sigma$
corresponding to the de Rham differential $d$ of $\Sigma$. (The choice of the
sign of the affine term is conventional.) We denote by
$\mathrm{Conn}(T[1]\Sigma,a_1{}^*Q)$
the affine space of generalized connections of $a_1{}^*Q$.

The adjoint and coadjoint bundles $\mathrm{Ad}\,a_1{}^*Q$,
$\mathrm{Ad}^\vee a_1{}^*Q$ are defined as
$\mathrm{Ad}\,a_1{}^*Q=a_1{}^*Q\times_G\mathfrak{g}$ and 
$\mathrm{Ad}^\vee a_1{}^*Q=a_1{}^*Q\times_G\mathfrak{g}^\vee$. 
A section $\bfs{s}\in\Gamma(T[1]\Sigma,\mathrm{Ad}\,a_1{}^*Q)$
is given locally on each open set $a_1{}^{-1}(U_A)$ of $T[1]\Sigma$
as a function $\bfs{s}_A\in\Gamma(a_1{}^{-1}(U_A),\mathfrak{g})$ with
$\bfs{s}_A=\mathrm{Ad}\,\bfs{g}_{AB}\,\bfs{s}_B$
on $a_1{}^{-1}(U_A)\cap a_1{}^{-1}(U_B)\not=\emptyset$
and similarly for $\mathrm{Ad}^\vee a_1{}^*Q$.
Degree shifting is achieved by replacing $\mathfrak{g}$
by $\mathfrak{g}[n]$ above and similarly for $\mathfrak{g}^\vee$.

The field content of the Weil sigma model is the following.

\begin{enumerate}

\item A section $\bfs{b}\in\Gamma(T[1]\Sigma,\mathrm{Ad}^\vee a_1{}^*Q[0])$.

\item A section $\bfs{B}\in\Gamma(T[1]\Sigma,\mathrm{Ad}^\vee a_1{}^*Q[-1])$.

\item A generalized connection $\bfs{c}\in\mathrm{Conn}(T[1]\Sigma,a_1{}^*Q)$.

\item A section $\bfs{C}\in\Gamma(T[1]\Sigma,\mathrm{Ad}\,a_1{}^*Q[2])$
\footnote{$\vphantom{\bigg[}$ In \cite{Zucchini6} $\bfs{B}$ $\bfs{C}$ were
denoted by $\mathrm{B}$, $\Gamma$, respectively.}.

\end{enumerate}

\noindent

The BV odd symplectic form is given by 
\begin{equation}
\Omega_W=\int_{T[1]\Sigma}\varrho\Big[\delta\bfs{b}_i\delta\bfs{c}^i
+\delta\bfs{B}_i\delta\bfs{C}^i\Big].
\label{OmegaW}
\end{equation}

The action of the Weil sigma model is given by
\begin{equation}
S_W=\int_{T[1]\Sigma}\varrho\Big[\bfs{b}_i\big(\bfs{d}\bfs{c}^i
-\hbox{$\frac{1}{2}$}f^i{}_{jk}\bfs{c}^j\bfs{c}^k+\bfs{C}^i\big)
-\bfs{B}_i\big(\bfs{d}\bfs{C}^i-f^i{}_{jk}\bfs{c}^j\bfs{C}^k\big)\Big].
\label{SW}
\end{equation}
$S_W$ satisfies the classical BV master equation
\begin{equation}
(S_W,S_W)_W=0,
\label{SWSW=0}
\end{equation}
where $(\cdot,\cdot)_W$ are the BV antibrackets associated with 
the BV form $\Omega_W$ \cite{Zucchini6}. 

The BV variations of the Weil sigma model fields are 
\begin{subequations}
\label{dWsuperfields}
\begin{align}
\delta_W\bfs{b}_i&=\bfs{d}\bfs{b}_i+f^k{}_{ji}\bfs{c}^j\bfs{b}_k+f^k{}_{ji}\bfs{C}^j\bfs{B}_k,
\label{dWbibfs}
\\
\delta_W\bfs{c}^i&=\bfs{d}\bfs{c}^i-\hbox{$\frac{1}{2}$}f^i{}_{jk}\bfs{c}^j\bfs{c}^k+\bfs{C}^i,
\label{dWcibfs}
\\
\delta_W\bfs{B}_i&=\bfs{d}\bfs{B}_i+f^k{}_{ji}\bfs{c}^j\bfs{B}_k-\bfs{b}_i,
\label{dWBibfs}
\\
\delta_W\bfs{C}^i&=\bfs{d}\bfs{C}^i-f^i{}_{jk}\bfs{c}^j\bfs{C}^k,
\label{dWCibfs}
\end{align}
\end{subequations}
where $\delta_W=(S_W,\cdot)_W$ \cite{Zucchini6}. 

From \eqref{SWSW=0}, it follows that the Weil sigma model action 
is BV invariant
\begin{equation}
\delta_WS_W=0.
\end{equation}
Again from \eqref{SWSW=0}, it follows that the Weil sigma model BV variation
operator $\delta_W$ is nilpotent \hphantom{xxxxxxxxxxxxxxxxxxxxxxxxxxxxx}
\begin{equation}
\delta_W{}^2=0,
\end{equation}
as can be directly verified from \eqref{dWsuperfields}.

\vskip .2cm
{\it Relation to the Weil algebra complex}.

The Weil sigma model owes its name to its relation to the Weil algebra complex 
of $\mathfrak{g}$, as we shall review next (see for instance \cite{Cartan1,Cartan2} 
for background material). 
To any Lie algebra $\mathfrak{g}$, there is canonically associated the Weil 
algebra $W(\mathfrak{g})=\wedge^*\mathfrak{g}^\vee[1]\otimes
\vee^*\mathfrak{g}^\vee[2]$, the tensor product of 
the antisymmetric and symmetric algebras of $\mathfrak{g}^\vee$
in degree $1$ and $2$, respectively. The natural $\mathfrak{g}$--valued
generators $\omega$, $\Omega$ of $W(\mathfrak{g})$ carry degrees $1$, $2$, 
respectively. The Weil operator $d_W$ is the degree $+1$ derivation 
on $W(\mathfrak{g})$ defined by
\begin{subequations}
\begin{align}
\label{}
d_W\omega^i&=\Omega^i-\frac{1}{2}f^i{}_{jk}\omega^j\omega^k,
\label{}
\\
d_W\Omega^i&=-f^i{}_{jk}\omega^j\Omega^k.
\label{}
\end{align}
\end{subequations}
It is simple to check that $d_W$ is nilpotent
\begin{equation}
d_W{}^2=0.
\label{dW2=0}
\end{equation}
Therefore, $(W(\mathfrak{g}),d_W)$ is a differential complex. 
Its cohomology $H^*(W(\mathfrak{g}),d_W)$ is actually trivial.
However, it is possible to define also a $\mathfrak{g}$ basic
cohomology $H^*_\mathrm{basic}(W(\mathfrak{g}),d_W)$, 
which turns out to be non trivial, as follows.
One defines degree $-1$ graded derivations $i_i$ and degree $0$ graded derivations
$l_i$ on $W(\mathfrak{g})$ by 
\begin{subequations}
\label{weildervs}
\begin{align}
i_i\omega^j&=\delta_i{}^j,
\label{iiomega}
\\
i_i\Omega^j&=0,
\label{iiOmega}
\\
l_i\omega^j&=-f^j{}_{ik}\omega^k,
\label{liomega}
\\
l_i\Omega^j&=-f^j{}_{ik}\Omega^k.
\label{liOmega}
\end{align}
\end{subequations}
The derivations $i_i$, $l_i$ and $d_W$ have the same formal properties
as the contraction $i_v$, Lie derivative $l_v$, with $v$ a vector field,
and de Rham differential $d_X$ on the graded algebra differential forms 
$\Omega^*(X)$ on a manifold $X$. The basic subalgebra
$W(\mathfrak{g})_\mathrm{basic}$ 
of $W(\mathfrak{g})$ consists of those elements $w \in W(\mathfrak{g})$ such that
\begin{subequations}
\label{weilbasic}
\begin{align}
i_iw&=0,
\label{}
\\
l_iw&=0. 
\label{}
\end{align}
\end{subequations}
$(W(\mathfrak{g})_\mathrm{basic},d_W)$ is a subcomplex of the differential complex 
$(W(\mathfrak{g}),d_W)$. Its cohomology $H^*(W(\mathfrak{g})_\mathrm{basic},d_W)$ is 
by definition the basic cohomology $H^*_\mathrm{basic}(W(\mathfrak{g}),d_W)$. 
It can be shown that $W(\mathfrak{g})_\mathrm{basic}=B\mathfrak{g}=\vee^*\mathfrak{g}^\vee[2]^G$ 
is the $G$--invariant subalge\-bra of the symmetric al\-ge\-bra 
$\vee^*\mathfrak{g}^\vee[2]\subset W(\mathfrak{g})$ of $\mathfrak{g}^\vee$ in degree $2$.
Actually, one has $H^*_\mathrm{basic}(W(\mathfrak{g}),d_W)\simeq B\mathfrak{g}$, 
since the restriction of $d_W$ to $B\mathfrak{g}$ vanishes
\footnote{$\vphantom{\bigg[}$ As is well known, the importance of the Weil
algebra basic cohomology stems from its being isomorphic to the 
cohomology of the classifying space $BG$ of $G$.}.

As shown in ref. \cite{Zucchini6}, when $Q$ is trivial, 
the superfields $\bfs{c}$, $\bfs{C}$ 
describe the embedding of $T[1]\Sigma$ into the Weil algebra 
$W(\mathfrak{g})$ of the Lie algebra $\mathfrak{g}$.
For any point $\bfs{z}\in T[1]\Sigma$,
the evaluation map $\mathrm{e}_{\bfs{z}}:\Gamma(T[1]\Sigma,W(\mathfrak{g}))
\mapsto W(\mathfrak{g})$ is a chain map of the chain complexes 
$(\Gamma(T[1]\Sigma,W(\mathfrak{g})),\delta'{}_W)$, $(W(\mathfrak{g}),d_W)$,
where $\delta'{}_W$ is the nilpotent mod $\bfs{d}$ reduction of $\delta_W$  
obtained by setting $\bfs{d}\bfs{c}^i=0$, $\bfs{d}\bfs{C}^i=0$ in \eqref{dWcibfs},
\eqref{dWCibfs}. When $Q$ is not trivial, $\bfs{c}$, $\bfs{C}$ become sections
of a vector bundle of Weil algebras. The above geometrical picture still holds but only
locally on $T[1]\Sigma$. This justifies the name given to the sigma model 
described here. 

We shall not attempt an exhaustive study of the BV cohomology of the Weil
sigma model. We shall only stress that it contains a sector isomorphic 
to the Weil algebra basic cohomology $H^*_\mathrm{basic}(W(\mathfrak{g}),d_W)$.
If one wished to construct a superfield out of a generic element 
$w\in W(\mathfrak{g})$, one would try with something like
\begin{equation}
\label{bfw}
\bfs{w}=\sum_{p,q}\frac{1}{p!q!}w_{i_1\ldots i_pj_1\ldots i_q}
\bfs{c}^{i_1}\cdots\bfs{c}^{i_p}\bfs{C}^{j_1} \cdots \bfs{C}^{j_q}.
\end{equation}
$\bfs{w}$, however, is only locally defined, since the superfields
$\bfs{c}^i$, $\bfs{C}^i$ are. To make $\bfs{w}$ globally defined, 
two requirements must be fulfilled. First, the right hand side of \eqref{bfw} 
must contain no occurrences of $\bfs{c}^i$,  since this is a generalized 
connection and, so, it is defined locally up to a local gauge
transformation. Second the Weil algebra element $w$ must be 
$G$--invariant. These requirements amount to requiring
$w\in B\mathfrak{g}$. So, we are led to consider superfields  
\begin{equation}
\label{bfwbg}
\bfs{w}=\sum_q\frac{1}{q!}w_{j_1\ldots i_q}\bfs{C}^{j_1} \cdots \bfs{C}^{j_q},
\end{equation}
with $w\in B\mathfrak{g}$. By a simple calculation, one finds 
\begin{equation}
\label{deltaw-dw}
\delta_W\bfs{w}=\bfs{d}\bfs{w}.
\end{equation}
Hence, $\bfs{w}$ is a cocycle of the mod $\bfs{d}$ BV cohomology. 
Since $H^*_\mathrm{basic}(W(\mathfrak{g}),d_W)\simeq B\mathfrak{g}$, 
the mapping $w\mapsto \bfs{w}$ defines an isomorphism of 
$H^*_\mathrm{basic}(W(\mathfrak{g}),d_W)$ and a certain sector of the 
mod $\bfs{d}$ BV cohomology.

In field theory, one is interested in the BV cohomology rather than the 
mod $\bfs{d}$ BV cohomology, since the BV cocycles are the observables of the 
field theory.
For any supercycle $\bfs{\mathcal{C}}$ of $T[1]\Sigma$ 
\begin{equation}
\label{<wC>}
\bfs{w}(\bfs{\mathcal{C}})=\oint_{\bfs{\mathcal{C}}}\bfs{w}
\end{equation}
is a cocycle of the BV cohomology
\begin{equation}
\label{}
\delta_W\bfs{w}(\bfs{\mathcal{C}})=0.
\end{equation}
For a fixed homology class $[\bfs{\mathcal{C}}]$ of $T[1]\Sigma$, 
the mapping $w\mapsto \bfs{w}(\bfs{\mathcal{C}})$ defines a
generally non injective
homomorphism of $H^*_\mathrm{basic}(W(\mathfrak{g}),d_W)$ into 
a certain sector of the BV cohomology.  

\vskip .2cm
{\it The Weil sigma model in components}.

One can expand the Weil sigma model fields in homogeneous components
\begin{subequations}
\label{bcBCi}
\begin{align}
\bfs{b}_i(\bfs{z})&=b_i(z)+\vartheta^\alpha A^+{}_{\alpha i}(z)
+\hbox{$\frac{1}{2}$}\vartheta^\alpha\vartheta^\beta c^+{}_{\alpha\beta i}(z),
\label{bi}
\\
\bfs{c}^i(\bfs{z})&=c^i(z)-\vartheta^\alpha A_\alpha{}^i(z)
-\hbox{$\frac{1}{2}$}\vartheta^\alpha\vartheta^\beta b^+{}_{\alpha\beta}{}^i(z),
\label{ci}
\\
\bfs{B}_i(\bfs{z})&=B_i(z)+\vartheta^\alpha \psi^+{}_{\alpha i}(z)
+\hbox{$\frac{1}{2}$}\vartheta^\alpha\vartheta^\beta C^+{}_{\alpha\beta i}(z),
\label{Bi}
\\
\bfs{C}^i(\bfs{z})&=C^i(z)-\vartheta^\alpha \psi_\alpha{}^i(z)
-\hbox{$\frac{1}{2}$}\vartheta^\alpha\vartheta^\beta B^+{}_{\alpha\beta}{}^i(z),
\label{Ci}
\end{align}
\end{subequations}
where $\bfs{z}\simeq(z,\vartheta)$, 
$z^\alpha$, $\vartheta^\alpha$ being base and fiber coordinates of $T[1]\Sigma$.
The ghost number of the various component fields is given by the degree of the
superfield they appear in minus the number of $\vartheta^\alpha$ they are
multiplied by, as $\mathrm{deg}\,\vartheta^\alpha=1$.
All component fields belong to either $\Omega^*(\Sigma,\mathrm{Ad}\,Q[n])$ or
$\Omega^*(\Sigma,\mathrm{Ad}^\vee Q[n])$ for some $n$
except for $A$ which is an ordinary connection of $Q$.
The choice of the signs of the component fields is conventional. 

The action and the BV variations of the Weil sigma model
(cf. eqs. \eqref{SW}, \eqref{dWsuperfields}) can be written down 
explicitly in terms of the components fields. The resulting expression are
lengthy and are collected in appendix \ref{sec:compweil} for convenience.

It is interesting to study the classical version of the Weil model and compare it 
with known models. 
The classical Weil sigma model is obtained by truncating the field content 
of the full Weil sigma model to the ghost number $0$ sector. 
The classical action of the model is found to be 
\begin{equation}
S_{Wc}=\int_\Sigma\Big[-b_i\big(F_A{}^i+B^{+i}\big)\Big].
\label{SWcl}
\end{equation}
This is essentially a $BF$ like field theory.
The symmetry variations of classical Weil sigma model are
obtained from the BV variations of the full Weil sigma model by retaining 
only the ghost fields of ghost number $1$,
\begin{subequations}
\label{dWcWeil}
\begin{align}
\delta_{Wc}A^i&=\psi^i-D_Ac^i,
\label{dWcAi}
\\
\delta_{Wc}b_i&=f^k{}_{ji}c^jb_k
\label{dWcbi}
\\
\delta_{Wc}B^{+i}&=D_A\psi^i-f^i{}_{jk}c^jB^{+k}
\label{dWcB+i}
\\
\delta_{Wc}c^i&=-\hbox{$\frac{1}{2}$}f^i{}_{jk}c^jc^k,\hskip .5cm
\label{dWcci}
\\
\delta_{Wc}\psi^i&=-f^i{}_{jk}c^j\psi^k.
\label{dWcpsii}
\end{align}
\end{subequations}
It is simple to verify that $S_{Wc}$ is invariant under the above 
field variations
\begin{equation}
\delta_{Wc}S_{Wc}=0.
\end{equation}
The classical field variation operator $\delta_{Wc}$ is nilpotent
\begin{equation}
\delta_{Wc}{}^2=0.
\end{equation}
We stress that this relation holds off--shell.

\vfill\eject

\begin{small}
\section{  \bf The gauge fixing of the Weil model}
\label{sec:yangmillsfixing}
\end{small}
To yield a field theory suitable for quantization, the Weil sigma model
has to be gauge fixed. To this end, we introduce two trivial pairs of fields
and their antifields.

\begin{enumerate}

\item $\tilde c\in \Omega^0(\Sigma,\mathrm{Ad}^\vee Q[-1])$,
$\gamma\in \Omega^0(\Sigma,\mathrm{Ad}^\vee Q[0])$ and their antifields

\noindent
$\tilde c^+\in \Omega^2(\Sigma,\mathrm{Ad}\,Q[0])$,
$\gamma^+\in \Omega^2(\Sigma,\mathrm{Ad}\,Q[-1])$. 

\item $\tilde C\in \Omega^0(\Sigma,\mathrm{Ad}^\vee Q[-2])$, 
$\Gamma\in \Omega^0(\Sigma,\mathrm{Ad}^\vee Q[-1])$ and their antifields

\noindent
$\tilde C^+\in \Omega^2(\Sigma,\mathrm{Ad}\,Q[1])$,
$\Gamma^+\in \Omega^2(\Sigma,\mathrm{Ad}\,Q[0])$. 

\end{enumerate} 

The Weil sigma model auxiliary BV odd symplectic form is
\begin{equation}
\Omega_{W\mathrm{aux}}=\int_\Sigma\Big[\delta\tilde c^{+i}\delta\tilde c_i
+\delta\gamma^{+i}\delta\gamma_i+\delta\tilde C^{+i}\delta\tilde C_i
+\delta\Gamma^{+i}\delta\Gamma_i\Big].
\end{equation}

The Weil sigma model auxiliary BV action is
\begin{equation}
\label{SWaux}
S_{W\mathrm{aux}}=\int_\Sigma\Big[\tilde c^{+i}\gamma_i+\tilde C^{+i}\Gamma_i\Big].
\end{equation}

The BV variations of the auxiliary fields are
\begin{subequations}
\label{dWaux}
\begin{align}
\delta_{W\mathrm{aux}}\tilde c_i&=\gamma_i,
\label{}
\\
\delta_{W\mathrm{aux}}\gamma_i&=0,
\label{}
\\
\delta_{W\mathrm{aux}}\gamma^{+i}&=-\tilde c^{+i},
\label{}
\\
\delta_{W\mathrm{aux}}\tilde c^{+i}&=0,
\label{}
\\
\delta_{W\mathrm{aux}}\tilde C_i&=-\Gamma_i,
\label{}
\\
\delta_{W\mathrm{aux}}\Gamma_i&=0,
\label{}
\\
\delta_{W\mathrm{aux}}\Gamma^{+i}&=-\tilde C^{+i},
\label{}
\\
\delta_{W\mathrm{aux}}\tilde C^{+i}&=0.
\label{}
\end{align}
\end{subequations}
One has as usual \hphantom{xxxxxxxxxxxxxxxxxx}
\begin{equation}
\delta_{W\mathrm{aux}}S_{W\mathrm{aux}}=0.
\end{equation}
$\delta_{W\mathrm{aux}}$ is nilpotent, \hphantom{xxxxxxxxxxxxxxxxxx}
\begin{equation}
\delta_{W\mathrm{aux}}{}^2=0.
\end{equation}

The gauge fixing is implemented by adding the auxiliary fields to the field
content of the Weil sigma model and by adding the auxiliary field action
$S_{W\mathrm{aux}}$ to the Weil sigma model action $S_W$: \hphantom{xxxxxxxxxxxxxxxxxx}
\begin{equation}
S_{W\mathrm{ext}}=S_W+S_{W\mathrm{aux}}.
\end{equation}
The gauge fixed action $I_W$ is obtained by restricting $S_{W\mathrm{ext}}$ to a
suitable Lagrangian submanifold $\mathfrak{L}_W$ in field space
\begin{equation}
I_W=S_{W\mathrm{ext}}\big|_{\mathfrak{L}_W}.
\label{I_Wdef}
\end{equation}
$I_W$ is invariant under a BRST symmetry $s_W$, which is the residual BV
symmetry left intact by the gauge fixing.

The Lagrangian submanifold $\mathfrak{L}_W$ is defined in terms of a ghost
number $-1$ gauge fermion $\Psi_W$ in the form $\phi^+=\delta
\Psi_W/\delta\phi$, where $\phi$ is any field. 
The gauge fermion we choose has the following form:
\begin{equation}
\Psi_W=\int_\Sigma\Big[-h^{ij}b_iB_j*1+\tilde C_iD_A*\psi^i+\tilde c_iD_{A_0}*(A^i-A_0{}^i)\Big],
\label{PsiW}
\end{equation}
where $h$ is an $\mathrm{Ad}$ invariant metric on $\mathfrak{g}$.
Above, $*$ denotes the Hodge operator associated with a metric of $\Sigma$. 
$A_0$ is a background connection of $Q$. The insertion of $A_0$ is
required by the global definedness on $\Sigma$ of the integrand in the right
hand side of \eqref{PsiW}.
Then, $\mathfrak{L}_W$ turns out to be explicitly defined by the 
constraints
\begin{subequations}
\label{LWdef}
\begin{align}
b^{+i}&=-h^{ij}B_j*1,
\label{}
\\
B^{+i}&=-h^{ij}b_j*1,
\label{}
\\
c^+{}_i&=0,
\label{}
\\
C^+{}_i&=0,
\label{}
\\
A^+{}_i&=*D_{A_0}\tilde c_i+f^k{}_{ji}*\psi^j\tilde C_k,
\label{}
\end{align}
\begin{align}
\psi^+{}_i&=-*D_A\tilde C_i,
\label{}
\\
\tilde C^{+i}&=D_A*\psi^i,
\label{}
\\
\Gamma^{+i}&=0,
\label{}
\\
\tilde c^{+i}&=D_{A_0}*(A^i-A_0{}^i),
\label{}
\\
\gamma^{+i}&=0.
\label{}
\end{align}
\end{subequations}

Substituting \eqref{LWdef} into \eqref{SWcomp} in accordance with \eqref{I_Wdef}, 
one then finds that the gauge fixed action $I_W$ is 
\begin{align}
I_W&=\int_\Sigma\Big[\gamma_iD_{A_0}*(A^i-A_0{}^i)+\tilde c_iD_{A_0}*\psi^i
-D_{A_0}\tilde c_i*D_Ac^i+h^{ij}b_ib_j*1
\label{I_W}
\\
&~~~~~~~~~~
-b_iF_A{}^i+B_iD_A\psi^i-D_A\tilde C_i*D_AC^i
-\big(\Gamma_i+f^j{}_{ki}\tilde  C_jc^k\big)D_A*\psi^i
\vphantom{\int_\Sigma}\nonumber
\\
&~~~~~~~~~~
+f^i{}_{jk}\tilde  C_i\psi^j*\psi^k
+f^{ij}{}_kB_iB_jC^k*1\Big].
\vphantom{\int_\Sigma}
\nonumber
\end{align}

The BRST variations of the fields are obtained from \eqref{dWWeil},
\eqref{dWaux} upon restriction to $\mathfrak{L}_W$. They read 
\begin{subequations}
\label{weilBRST}
\begin{align}
s_WA^i&=\psi^i-D_Ac^i,
\label{}
\\
s_W\psi^i&=-D_AC^i-f^i{}_{jk}c^j\psi^k,
\label{}
\\
s_Wb_i&=f^k{}_{ji}c^jb_k+f^k{}_{ji}C^jB_k,
\label{}
\\
s_WB_i&=-b_i+f^k{}_{ji}c^jB_k,
\label{}
\\
s_Wc^i&=C^i-\hbox{$\frac{1}{2}$}f^i{}_{jk}c^jc^k,
\label{}
\\
s_WC^i&=-f^i{}_{jk}c^jC^k,
\label{}
\\
s_W\tilde c^i&=\gamma^i,
\label{}
\\
s_W\gamma^i&=0,
\label{}
\\
s_W\tilde C^i&=-\Gamma^i,
\label{}
\\
s_W\Gamma^i&=0.
\label{}
\end{align}
\end{subequations}
One can verify directly that \hphantom{xxxxxxxxxxxxxxxxxxxxxxx}
\begin{equation}
s_WI_W=0.
\label{IW}
\end{equation}
Further, one has \hphantom{xxxxxxxxxxxxxxxxxxxxxxx}
\begin{equation}
s_W{}^2=0.
\label{sW2=0}
\end{equation}
In general, the BRST variation operator is nilpotent only on--shell. In this
case however, it does square to $0$ off--shell.

Using \eqref{weilBRST}, it can be verified that 
\begin{equation}
I_W=S_{W\mathrm{top}}+s_W\Psi_W,
\label{IW=SWtop+sWPsi}
\end{equation}
where the topological action $S_{W\mathrm{top}}$ is given by
\begin{equation}
S_{W\mathrm{top}}=\int_\Sigma\Big[-b_iF_A{}^i+B_iD_A\psi^i\Big].
\label{SWtop}
\end{equation}
This relation shows the topological nature of the theory.
All dependence on the metric of $\Sigma$ and the background connection $A_0$ 
is buried inside the gauge fermion $\Psi_W$. The topological quantum field 
correlators therefore are going to be independent from these data. 

The topological field theory, which we are dealing with, is in fact the 
$2$--dimensional version of Donaldson--Witten theory \cite{Witten5,Witten6}, which 
describes the moduli space of flat connection of a 
trivial principal $G$--bundle $Q$.
This is easily seen from the BRST variations 
\eqref{weilBRST} obtained above. It is known that a topological field theory 
localizes on the BRST invariant purely bosonic on--shell configurations. 
Setting all the fermionic fields to zero in the BRST variations and imposing 
that the resulting expressions vanish leads to the equation $b_i=0$, 
which, on shell, is equivalent to \hphantom{xxxxxxxxxxxxx}
\begin{equation}
F^i=0.
\label{Fi=0}
\end{equation}

We remark that the above procedure yields at once the full topological 
field theory action and the Faddeev--Popov gauge fixing action. The latter 
consists of those terms in the right hand side of \eqref{I_W}, which depend 
explicitly on the background connection $A_0$.

\vfill\eject

\begin{small}
\section{  \bf The Poisson--Weil sigma model}
\label{sec:gaugedpoisson}
\end{small}

The Poisson--Weil sigma model stems from coupling the Weil sigma model described in sect.
\ref{sec:yangmills} and the Poisson sigma model \cite{Ikeda2,Strobl}. This procedure is 
in fact a way of gauging the latter and generalizes our original construction
in \cite{Zucchini6}.  As for the Weil sigma model, the covariance of the
superfields of the version of the Poisson--Weil sigma model illustrated below is 
more general than that originally envisaged in  \cite{Zucchini6}, but its BV
formulation is essentially the same. The reader is therefore invited again to read that paper 
for more details on the BV formalism used and the derivation of the classical 
master equation and BV variations below. 

We consider a geometrical setting consisting of the following elements.

\vskip.35cm

\begin{enumerate}

\item A closed surface $\Sigma$.

\item A compact connected Lie group $G$ with Lie algebra $\mathfrak{g}$.

\item A principal $G$--bundle $Q$ over $\Sigma$.

\item A manifold $M$ carrying a smooth effective left $G$--action with 
fundamental vector field $u\in C^\infty(M, TM\otimes \mathfrak{g}^\vee)$ .

\item A $G$--invariant $2$--vector $P\in C^\infty(M,\wedge^2 TM)$ and a  
$G$--equivariant $\mathfrak{g}^\vee$--valued scalar $\mu\in C^\infty(M,\mathfrak{g}^\vee)$. 

\end{enumerate}

The geometry associated with these data is rich and intricate.
Some of its features have already emerged in the work \cite{Baptista1,Baptista2,Baptista3}.
Here we shall limit ourselves to indicate the aspects of it which are most
directly relevant in our analysis.

The first three geometrical data are the ones entering in the definition of
the Weil sigma model as illustrated in sect. \ref{sec:yangmills}. The fourth geometrical
datum allows one to define the bundle $E_M=Q\times_G M$ with base $\Sigma$. 
$E_M$ can be described as follows. 
Let $\{U_A\}$ be a sufficiently fine open covering of $\Sigma$. Then,  
locally, one has $E_M|_{U_A}\simeq U_A\times M$. 
$E_M$ is obtained by identifying $(z,m_A)\in U_A\times M$ and $(z,m_B)\in U_B\times M$ with
$z\in U_A\cap U_B\not=\emptyset$ and $m_A=g_{AB}(z)(m_B)$, where 
$\{g_{AB}\}$ is the $G$--valued $1$--cocycle representing $Q$ with respect to 
$\{U_A\}$ and $g(m)$ denotes the action of the group element $g\in G$ on the
point $m\in M$. When $Q$ is trivial, one has $E_M\simeq \Sigma\times M$.  
Sections $x\in\Gamma(\Sigma, E_M)$ generalize the customary embeddings
$x:\Sigma\rightarrow M$, which they reduce to when $Q$ is trivial. 

The bundle projection $a_1:T[1]\Sigma\rightarrow \Sigma$ introduced in sect.  
\ref{sec:yangmills} allows one to pull-back $E_M$ to $T[1]\Sigma$ yielding the
bundle $a_1{}^*E_M$ with base space $T[1]\Sigma$. In terms of a fine open 
covering $\{U_A\}$ of $\Sigma$,
one has $a_1{}^*E_M|_{a_1{}^{-1}(U_A)}\simeq a_1{}^{-1}(U_A)\times M$. 
$a_1{}^*E_M$ is obtained by identifying 
$(\bfs{z},m_A)\in a_1{}^{-1}(U_A)\times M$ and $(\bfs{z},m_B)\in a_1{}^{-1}(U_B)\times M$ with
$a_1(\bfs{z})\in U_A\cup U_B\not=\emptyset$ and $m_A=\bfs{g}_{AB}(\bfs{z})(m_B)$, where 
$\{\bfs{g}_{AB}\}$ is the $G$--valued $1$--cocycle representing $a_1{}^*Q$ with respect to 
$\{a_1{}^{-1}(U_A)\}$ defined in sect.  \ref{sec:yangmills}.
When $Q$ is trivial, one has $a_1{}^*E_M\simeq T[1]\Sigma\times M$.  
Sections $\bfs{x}\in\Gamma(T[1]\Sigma, a_1{}^*E_M)$ generalize the customary superembeddings
$x:T[1]\Sigma\rightarrow M$, which they reduce to when $Q$ is trivial. 

Associated with $E_M$ are the vector bundle
$\mathrm{Vert}\,TE_M$ and its dual 
$\mathrm{Vert}^*TE_M$ with base space $E_M$, where 
$\mathrm{Vert}\,TE_M=\mathrm{ker}\,\pi_{E_M*}$, $\pi_{E_M}:E_M\rightarrow \Sigma$
being the bundle projection and $\pi_{E_M*}:TE_M\rightarrow T\Sigma$ its tangent
map. Given a fine enough open covering $\{U_A\}$ of $\Sigma$,
the transition functions of the bundle $\mathrm{Vert}\,TE_M$ with respect to
the open covering $\{\pi_{E_M}{}^{-1}(U_A)\}$ are of the form $t_{AB}(e)=g_{AB}(z)_*(m_B)$
for $\pi_{E_M}(e)\in U_A\cap U_B$, 
where $e\simeq (z,m_B)$ in the trivialization $E_M|_{U_B}\simeq U_B\times M$
and  $g_*(m):T_mM\rightarrow T_{g(m)}M$ is the tangent map at $m\in M$ of
the action $g:M\rightarrow M$ of $g\in G$.  
Given $x\in \Gamma(\Sigma, E_M)$, one can define the pull--back bundles 
$x^*\mathrm{Vert}\,TE_M$ and $x^*\mathrm{Vert}^*TE_M$, which are vector
bundles with base space $\Sigma$. The transition functions of the bundle
$x^*\mathrm{Vert}\,TE_M$ are $t_{AB}(z)=g_{AB}(z)_*(x_B(z))$,
where $x(z)\simeq (z,x_B(z))$ in the trivialization $E_M|_{U_B}\simeq U_B\times M$.

This construction can be extended by replacing $E_M$ by $a_1{}^*E_M$ and 
$\Sigma$ by $T[1]\Sigma$ throughout above. In this way, one  
builds the vector bundles $\mathrm{Vert}\,Ta_1{}^*E_M$ and its dual 
$\mathrm{Vert}^*Ta_1{}^*E_M$ with base $a_1{}^*E_M$. The transition function
of $\mathrm{Vert}\,Ta_1{}^*E_M$ are of the form $\bfs{t}_{AB}(\bfs{e})=\bfs{g}_{AB}(\bfs{z})_*(m_B)$
for $\pi_{a_1{}^*E_M}(\bfs{e})\in a_1{}^{-1}(U_A)\cap a_1{}^{-1}(U_B)$,
where $\bfs{e}\simeq (\bfs{z},m_B)$ in the trivialization 
$a_1{}^*E_M|_{a_1{}^{-1}(U_B)}\simeq a_1{}^{-1}(U_B)\times M$.
Given $\bfs{x}\in \Gamma(T[1]\Sigma, a_1{}^*E_M)$, one can build the pull--back  bundles
$\bfs{x}^*\mathrm{Vert}\,Ta_1{}^*E_M$ and $\bfs{x}^*\mathrm{Vert}^*Ta_1{}^*E_M$, 
which are vector bundles with base space $T[1]\Sigma$. 
The transition functions of the bundle
$\bfs{x}^*\mathrm{Vert}\,Ta_1{}^*E_M$ are 
$\bfs{t}_{AB}(\bfs{z})=\bfs{g}_{AB}(\bfs{z})_*(\bfs{x}_B(\bfs{z}))$,
where $\bfs{x}(\bfs{z})\simeq (\bfs{z},\bfs{x}_B(\bfs{z}))$ in 
the trivialization $a_1{}^*E_M|_{a_1{}^{-1}(U_B)}\simeq a_1{}^{-1}(U_B)\times M$.


The fundamental vector field $u$ satisfies the basic equivariance relation
\footnote{$\vphantom{\bigg[}$ This relation fixes also the overall sign
convention for $u$ used in the paper. According to this,
$g^a(m)=m^a-\xi^iu_i{}^a(m)+O(\xi^2)$
for $g=\exp(\xi)\in G$ with $\xi\in\mathfrak{g}$. See appendix \ref{sec:covariance}.}
\begin{equation}
[u_i,u_j]^a=u_i{}^b\partial_bu_j{}^a-u_j{}^b\partial_bu_i{}^a=f^k{}_{ij}u_k{}^a.
\end{equation}
The $G$--invariance of the $2$--vector $P$ and the $G$--equivariance of the scalar $\mu$ 
are crucial in our construction. Infinitesimally, they are equivalent to 
the relations
\begin{subequations}
\label{Gvariance}
\begin{align}
l_{u_i}P^{ab}&=u_i{}^c\partial_cP^{ab}-\partial_cu_i{}^aP^{cb}-\partial_cu_i{}^bP^{ac}=0.
\label{luiP=0}
\\
l_{u_i}\mu_j&=u_i{}^b\partial_b\mu_j=f^k{}_{ij}\mu_k,
\label{luimuj=fkijmuk}
\end{align}
\end{subequations}

The field content of the Poisson--Weil sigma model consists the following superfields.

\begin{enumerate}

\item The superfields of the Weil sigma model.

\item A section $\bfs{x}\in\Gamma(T[1]\Sigma,a_1{}^*E_M)$.

\item A section $\bfs{y}\in\Gamma(T[1]\Sigma,\bfs{x}^*\mathrm{Vert}^* Ta_1{}^*E_M[1])$.

\end{enumerate}

The BV odd symplectic form is given by 
\begin{equation}
\label{OmegaPW}
\Omega_{PW}=\Omega_W+\int_{T[1]\Sigma}\varrho\delta\bfs{x}^a\delta\bfs{y}_a,
\end{equation}
where $\Omega_W$ is the  BV odd symplectic form of the Weil sigma model given in 
\eqref{OmegaW}. 

The action of the Poisson--Weil sigma model is
\begin{equation}
S_{PW}=S_W+\int_{T[1]\Sigma}\varrho\Big[
\bfs{y}_a\big(\bfs{d}\bfs{x}^a+u_i{}^a(\bfs{x})\bfs{c}^i\big)+\mu_i(\bfs{x})\bfs{C}^i
+\hbox{$\frac{1}{2}$}P^{ab}(\bfs{x})\bfs{y}_a\bfs{y}_b\Big],
\label{SPW}
\end{equation}
where $S_W$ is the action of the Weil sigma model given in \eqref{SW}.
The $G$--invariance of $P$ and the $G$--equivariance of $\mu$ 
ensure the global definedness of the integrand in the right
hand side of \eqref{SPW}. 

It can be verified by explicit computation that $S_{PW}$ satisfies the
classical master equation \hphantom{xxxxxxxxxxxxxxxxxxxxxx}
\begin{equation}
(S_{PW},S_{PW})_{PW}=0,
\label{SWPSWPWP=0}
\end{equation}
where $(\cdot,\cdot)_{PW}$ are the BV antibrackets associated with 
the BV form $\Omega_{PW}$, provided $u$, $\mu$, $P$ satisfy the conditions
\begin{subequations}
\label{BVconds}
\begin{align}
&P^{ad}\partial_dP^{bc}+P^{bd}\partial_dP^{ca}+P^{cd}\partial_dP^{ab}=0,
\label{Aabc=0}
\\
&u_i{}^a+P^{ab}\partial_b\mu_i=0,
\label{Si=0}
\end{align}
\end{subequations}
\cite{Zucchini6}. \eqref{Aabc=0} \eqref{Si=0} imply respectively the following properties.

\begin{enumerate}

\item $P$ is a Poisson 2--vector and $M$ is thus a Poisson manifold. 

\item $\mu$ is a moment map for the $G$--action, which is therefore Hamiltonian.

\end{enumerate}

\noindent

The BV variations of the Poisson--Weil sigma model fields are
\begin{subequations}
\label{dPWsuperfields}
\begin{align}
\delta_{PW}\bfs{b}_i&=\delta_W\bfs{b}_i-u_i{}^a(\bfs{x})\bfs{y}_a,
\label{dWPbibfs}
\\
\delta_{PW}\bfs{c}^i&=\delta_W\bfs{c}^i,
\label{dWPcibfs}
\\
\delta_{PW}\bfs{B}_i&=\delta_W\bfs{B}_i-\mu_i(\bfs{x}),
\label{dWPBibfs}
\\
\delta_{PW}\bfs{C}^i&=\delta_W\bfs{C}^i,
\label{dWPCibfs}
\\
\delta_{PW}\bfs{x}^a&=\bfs{d}\bfs{x}^a+u_i{}^a(\bfs{x})\bfs{c}^i+P^{ab}(\bfs{x})\bfs{y}_b,
\label{dWPxabfs}
\\
\delta_{PW}\bfs{y}_a&=\bfs{d}\bfs{y}_a+\partial_au_i{}^b(\bfs{x})\bfs{y}_b\bfs{c}^i
+\partial_a\mu_i(\bfs{x})\bfs{C}^i
+\hbox{$\frac{1}{2}$}\partial_aP^{bc}(\bfs{x})\bfs{y}_b\bfs{y}_c,
\label{dWPyabfs}
\end{align}
\end{subequations}
where $\delta_{PW}=(S_{PW},\cdot)_{PW}$ and  
the Weil sigma model $\delta_W$ variations are given by \eqref{dWsuperfields} \cite{Zucchini6}.

From \eqref{SWPSWPWP=0}, it follows that the Poisson--Weil sigma model action 
is BV invariant \hphantom{xxxxxxxxxxxxxxxxxxxxxxx}
\begin{equation}
\delta_{PW}S_{PW}=0.
\end{equation}
Again from \eqref{SWPSWPWP=0}, it follows that the Poisson--Weil sigma model BV variation
operator $\delta_{PW}$ is nilpotent \hphantom{xxxxxxxxxxxxxxxxxxxxxxx}
\begin{equation}
\delta_{PW}{}^2=0.
\end{equation}

\vskip .2cm
{\it Relation to the Hamiltonian basic Poisson--Lichnerowicz cohomology}.

When conditions \eqref{BVconds}, \eqref{Gvariance} are satisfied,
if $a\in\mathfrak{g}^\vee$ with coadjoint orbit $\mathcal{O}_a$ and 
$\mu^{-1}(\mathcal{O}_a)$ is a submanifold of $M$ on which $G$ acts freely
and properly, then the quotient $M_a=\mu^{-1}(\mathcal{O}_a)/G$ inherits a 
Poisson structure $P_a$, by a classic result of Marsden and Ratiu \cite{Ratiu1}. 
One considers mainly $M_0=\mu^{-1}(\{0\})/G\equiv M/\!/G$.

From the above discussion, it appears that the Poisson--Weil sigma model on a
Poisson manifold $M$ carrying a Hamiltonian action of a group $G$ encodes the
Hamiltonian  reduction of $M$ by $G$. Upon suitably restricting 
the image of $\bfs{x}$ to $\mu^{-1}(\mathcal{O}_a)$, one expects to
obtain some kind of 
sigma model on $M_a$. When the principal bundle $Q$ is trivial, 
this should be an ordinary Poisson sigma model on $M_a$. 
The embedding fields of the model are then just maps $x:\Sigma \rightarrow M_a$.
Conversely, when $Q$ is non trivial, one should obtain a generalized Poisson 
sigma model on $M_a$. The embedding fields of the model are then sections 
$x\in \Gamma(\Sigma,E_M)$ such that, in any trivialization $E_M|_{U_A}\simeq U_A\times M$, 
$x_A(z)\in \mu^{-1}(\mathcal{O}_a)$ for $z\in U_A$.  
(Note that this property is independent from the chosen trivialization). 
Intuitively, they are some kind of ``$Q$--twisted" maps $x:\Sigma \rightarrow M_a$.
These facts should be reflected in the BV cohomology of the Poisson--Weil sigma
model, which we explore next. As we shall see, this investigation will bring
us close to the boundary of known mathematics.

Recall that a Poisson manifold $M$ with Poisson $2$--vector field $P$ is characterized by the algebra
of multivector fields $C^\infty(M,\wedge^*TM)$ and by the Poisson--Lichne\-ro\-wicz 
differential $\sigma_\mathrm{PL}=[P,\cdot]$, where $[\cdot,\cdot]$ are the Schouten brackets
on $C^\infty(M,\wedge^*TM)$ (see for instance \cite{Vaisman} for background material).
Since $\sigma_\mathrm{PL}$ is nilpotent, $(C^\infty(M,\wedge^*TM),\sigma_\mathrm{PL})$
is a differential complex, the Poisson--Lichnerowicz complex.
The associated cohomology is the Poisson--Lichnerowicz cohomology
$H^*_\mathrm{PL}(M)$. Each cohomology class is represented 
by a Poisson--Lichnerowicz cocycle, that is 
a multivector field $\alpha\in C^\infty(M,\wedge^*TM)$ 
satisfying 
\begin{equation}
\sigma_\mathrm{PL}\alpha=0.
\label{sigmaPLalpha=0}
\end{equation}
This cocycle is defined up to a Poisson--Lichnerowicz coboundary, i. e.
a multivector field belonging to the image of $\sigma_\mathrm{PL}$.

A simple analysis shows that $H^0_\mathrm{PL}(M)$ is the algebra of Casimir functions
of $M$ and $H^1_\mathrm{PL}(M)$ is the quotient of the space of Poisson vector
fields of $M$ over the space of Hamiltonian vector fields, etc. Further, 
the Poisson $2$--vector field $P$, viewed as an element 
of $C^\infty(M,\mathrm{Hom}(T^*M,TM))$, induces a homomorphism 
$P^\#$ of the ordinary de Rham cohomology $H^*_\mathrm{dR}(M)$ into
$H^*_\mathrm{PL}(M)$, which is an isomorphism in the symplectic case. 
 
Suppose that $M$ carries a Hamiltonian smooth effective left $G$--action with 
fundamental vector field $u\in C^\infty(M, TM\otimes \mathfrak{g}^\vee)$ and 
$G$--equivariant moment map $\mu\in C^\infty(M,\mathfrak{g}^\vee)$
and leaving $P$ invariant.
We call a multivector field $\alpha\in C^\infty(M,\wedge^*TM)$ Hamiltonian
basic, if $\alpha$ sa\-tis\-fies the conditions 
\begin{subequations}
\label{poissonbasic}
\begin{align}
&i_{d\mu_i}\alpha=0,
\label{idmuialpha=0}
\\
&l_{u_i}\alpha=0,
\label{luialpha=0}
\end{align}
\end{subequations}
where $i_\omega$ denotes contraction with the $1$--form
$\omega\in C^\infty(M,T^*M)$ and $l_v$ is the Lie derivative along the vector 
field $v\in C^\infty(M,TM)$, i. e. if $\alpha$ is $G$--invariant and tangent 
to the $\mu$ fibers. The terminology is justified by the analogy to the notion 
of basic forms of a manifold with a group action. 
We denote by $C^\infty(M,\wedge^*TM)_\mathrm{basic}$ the
subalgebra of $C^\infty(M,\wedge^*TM)$ spanned by the Hamiltonian
basic multivector fields. Using the relations
\begin{subequations}
\begin{align}
&i_{d\mu_i}\sigma_\mathrm{PL}+\sigma_\mathrm{PL}i_{d\mu_i}=l_{u_i},
\label{}
\\
&\sigma_\mathrm{PL}l_{u_i}-l_{u_i}\sigma_\mathrm{PL}=0,
\label{}
\\
&l_{u_i}i_{d\mu_i}-l_{u_i}i_{d\mu_i}=f^k{}_{ij}i_{d\mu_k},
\label{}
\end{align}
\end{subequations}
one shows that $(C^\infty(M,\wedge^*TM)_\mathrm{basic},\sigma_\mathrm{PL})$
is a subcomplex of $(C^\infty(M,\wedge^*TM)$, $\sigma_\mathrm{PL})$, the Hamiltonian basic 
PL complex. The associated cohomology is the Hamiltonian
basic Poisson--Lichnerowicz cohomology $H^*_\mathrm{PLbasic}(M)$. 
Each cohomology class is represented by a Hamiltonian
basic Poisson--Lichnerowicz cocycle, i. e. 
a multivector field $\alpha\in C^\infty(M,\wedge^*TM)$ satisfying the conditions
\eqref{sigmaPLalpha=0}, \eqref{poissonbasic}.
This cocycle is defined up to a Hamiltonian
basic Poisson--Lichnerowicz coboundary, i. e.
a multivector field belonging to the image of $\sigma_\mathrm{PL}$
restricted to $C^\infty(M,\wedge^*TM)_\mathrm{basic}$.

Repeating the analysis done for ordinary Poisson cohomology, one can show that
$H^0_\mathrm{PLbasic}(M)$ is the algebra of ordinary Casimir functions
of $M$ and $H^1_\mathrm{PLbasic}(M)$ is the quotient of the space of
$G$--invariant Poisson vector fields of $M$ tangent to the $\mu$ fibers over the 
space of Hamiltonian vector fields with $G$--invariant Hamiltonians, etc.
Further, $P$ induces a homomorphism 
$P^\#$ of the ordinary basic de Rham cohomology $H^*_\mathrm{dRbasic}(M)$ into
$H^*_\mathrm{PLbasic}(M)$, which is an isomorphism in the symplectic case. 

The Hamiltonian basic Poisson--Lichnerowicz cohomology $H^*_\mathrm{PLbasic}(M)$
was introduced and studied in a more general context by Ginzburg
in \cite {Ginzburg}. It is natural to expect $H^*_\mathrm{PLbasic}(M)$
to be related to the Poisson--Lichnerowicz cohomology of
the reduced Poisson manifolds $M_a$ defined above. However, to the best of our
knowledge, so far this relation has not been elucidated in the mathematical
literature except for symplectic manifolds in \cite{Kirwan} by Kirwan, 
who showed the existence of a natural surjective generally non injective homomorphism
$\kappa:H^*_\mathrm{dRbasic}(M) \simeq H^*_\mathrm{PLbasic}(M)\rightarrow 
H^*_\mathrm{dR}(M_0)$. Virtually nothing is known for more general Poisson manifolds. 

We shall not attempt an exhaustive study of the BV cohomology of the Poisson--Weil
sigma model. We shall only try to highlight some of its novel features and its
relation to the Hamiltonian basic Poisson--Lichnerowicz cohomology.
If one wished to construct a superfield out of a generic multivector 
field $\alpha\in C^\infty(M,\wedge^*TM)$, one would start by trying  with something 
of the form 
\begin{equation}
\bfs{\alpha}=\sum_p\frac{1}{p!}\alpha^{a_1\ldots a_p}(\bfs{x})\bfs{y}_{a_1}\cdots\bfs{y}_{a_p}.
\label{bfsalpha}
\end{equation}
This object is however only locally defined since the superfields $\bfs{x}^a$,
$\bfs{y}_a$ are defined only up to a local $G$--action. 
To render $\bfs{\alpha}$ globally defined, one has to demand
the multivector field $\alpha$ to be $G$--invariant, $\alpha\in C^\infty(M,\wedge^*TM)^G$. 
Infinitesimally, this is equivalent to \eqref{luialpha=0}. 

A straightforward calculation yields
\begin{align}
\delta_{PW}\bfs{\alpha}&=\bfs{d}\bfs{\alpha}
+\sum_p\frac{1}{p!}(i_{d\mu_i}\alpha)^{a_1\ldots a_p}(\bfs{x})
\bfs{C}^i\bfs{y}_{a_1}\cdots\bfs{y}_{a_p}
\label{}
\\
&\hskip 4cm
-\sum_p\frac{1}{p!}(\sigma_\mathrm{PL}\alpha)^{a_1\ldots a_p}
(\bfs{x})\bfs{y}_{a_1}\cdots\bfs{y}_{a_p}.
\nonumber
\end{align}
Hence, one has \hphantom{xxxxxxxxxxxxxxxxxxxxxxxxxxxxxxxxxxxxxxxxxxxx}
\begin{equation}
\delta_{PW}\bfs{\alpha}=\bfs{d}\bfs{\alpha}
\label{BVmodd}
\end{equation}
provided $\alpha\in C^\infty(M,\wedge^*TM)^G$ satisfies 
\eqref{sigmaPLalpha=0}, \eqref{idmuialpha=0}.
In that case, $\bfs{\alpha}$ is a cocycle of the mod $\bfs{d}$ BV cohomology. 
Furthermore, the mapping $\alpha\mapsto \bfs{\alpha}$ defines an isomorphism of 
$H^*_\mathrm{PLbasic}(M)$ and a distinguished sector of the 
mod $\bfs{d}$ BV cohomology.

As already remarked earlier, in 
field theory, one is interested in the BV cohomology rather than the 
mod $\bfs{d}$ BV cohomology, since BV cocycles are observables.
For any supercycle $\bfs{\mathcal{C}}$ of $T[1]\Sigma$, \hphantom{xxxxxxx}
\begin{equation}
\label{<PWwC>}
\bfs{\alpha}(\bfs{\mathcal{C}})=\oint_{\bfs{\mathcal{C}}}\bfs{\alpha}
\end{equation}
is a cocycle of the BV cohomology
\begin{equation}
\label{}
\delta_{PW}\bfs{\alpha}(\bfs{\mathcal{C}})=0.
\end{equation}
For a fixed homology class $[\bfs{\mathcal{C}}]$ of $T[1]\Sigma$, 
the mapping $\alpha\mapsto \bfs{\alpha}(\bfs{\mathcal{C}})$ defines a
generally non injective homomorphism of $H^*_\mathrm{PLbasic}(M)$ into 
the BV cohomology. 

We conclude that, for a fixed homology class
$[\bfs{\mathcal{C}}]$ of $T[1]\Sigma$, 
$\bfs{\alpha}(\bfs{\mathcal{C}})$ is an observable provided $\alpha$ is 
a Hamiltonian basic Poisson--Lichnerowicz cocycle. This establishes a homomorphism 
of the Hamiltonian basic Poisson--Lichnerowicz cohomology
$H^*_\mathrm{PLbasic}(M)$ into the Poisson--Weil BV cohomology.

In \cite {Ginzburg}, Ginzburg also defined the equivariant Poisson--Lichnerowicz cohomology 
$H^*_{\mathrm{PL}G}(M)$. This can be realized in two different but equivalent models. 
In the Weil model, one relies on the Weil algebra $(W(\mathfrak{g}),d_W)$
complex described in sect. \ref{sec:yangmills}. 
$H^*_{\mathrm{PL}G}(M)$ is the cohomology of the complex 
$((C^\infty(M,\wedge^*TM)\otimes W(\mathfrak{g}))_\mathrm{basic},\sigma_{PLW})$,
where basicity is defined in terms of the graded derivations
$i_{Wi}=i_{d\mu_i}+i_i$, $l_{Wi}=l_{u_i}+l_i$, by extending \eqref{weilbasic}, \eqref{poissonbasic} 
in obvious fashion, and $\sigma_{PLW}=\sigma_{PL}+d_W$.
In the Cartan model, $H^*_{\mathrm{PL}G}(M)$ is the cohomology of the complex 
$((C^\infty(M,\wedge^*TM)\otimes \vee^*\mathfrak{g}^\vee
[2])^G,\sigma_{PLC})$, where $G$--invariance is defined in terms of 
$l_{Ci}=l_{u_i}+l_i$ and $\sigma_{PLC}=\sigma_{PL}-\Omega^ki_{d\mu_k}$, 
with $\Omega^i$ the degree $2$ generators of $\vee^*\mathfrak{g}^\vee
[2]$ and $l_i$ defined as in \eqref{liOmega}. 

When $G$ is compact and $\mu$ is a submersion onto $\mathfrak{g}^\vee$, 
$H^*_\mathrm{PLbasic}(M)$ is isomorphic to the equivariant Poisson--Lichnerowicz cohomology  
$H^*_{\mathrm{PL}G}(M)$ \cite{Ginzburg}. Using the Cartan model for simplicity, 
a class of $H^*_\mathrm{PLbasic}(M)$ is represented by a $G$--invariant multivector field 
$\alpha\in (C^\infty(M,\wedge^*TM)\otimes \vee^*\mathfrak{g}^\vee [2])^G$,
\begin{equation}
l_{Ci}\alpha=0,
\end{equation}
satisfying the cocycle condition 
\begin{equation}
\sigma_{PLC}\alpha=0.
\label{sigmaPLCalpha=0}
\end{equation}
This suggests a possible generalization of the ansatz \eqref{bfsalpha} of the
form
\begin{equation}
\bfs{\alpha}=\sum_{p,q}\frac{1}{p!q!}\alpha^{a_1\ldots a_p}{}_{i_1\ldots i_q}(\bfs{x})
\bfs{y}_{a_1}\cdots\bfs{y}_{a_p}\bfs{C}^{i_1}\cdots\bfs{C}^{i_q},
\label{bfsalphaCC}
\end{equation}
where $\alpha\in (C^\infty(M,\wedge^*TM)\otimes \vee^*\mathfrak{g}^\vee [2])^G$. 
The $G$--invariance of $\alpha$ is required by the proper global definedness of the
superfield $\bfs{\alpha}$. A straightforward calculation leads to
\begin{equation}
\delta_{PW}\bfs{\alpha}=\bfs{d}\bfs{\alpha}
-\sum_{p,q}\frac{1}{p!q!}(\sigma_\mathrm{PLC}\alpha)^{a_1\ldots a_p}{}_{i_1\ldots i_q}
(\bfs{x})\bfs{y}_{a_1}\cdots\bfs{y}_{a_p}\bfs{C}^{i_1}\cdots\bfs{C}^{i_q}.
\nonumber
\end{equation}
Hence, $\bfs{\alpha}$ satisfies \eqref{BVmodd}, 
provided $\alpha\in C^\infty(M,\wedge^*TM)^G$ satisfies 
\eqref{sigmaPLCalpha=0}. In this way, proceeding exactly in the same way as
above, one can construct observables of the field theory. However, this procedure is not going to
yield genuinely new observables. In fact, the inclusion 
$C^\infty(M,\wedge^*TM)_\mathrm{basic}\subset 
(C^\infty(M,\wedge^*TM)\otimes \vee^*\mathfrak{g}^\vee [2])^G$ induces the
isomorphism $H^*_\mathrm{PLbasic}(M)\simeq H^*_{\mathrm{PL}G}(M)$ mentioned above.
This means that any mod $\bfs{d}$ BV cocycle of the form \eqref{bfsalphaCC}
is always BV cohomologous to one of the form \eqref{bfsalpha}.

\vskip .2cm
{\it The Poisson--Weil sigma model in components}.

One can expand the Poisson--Weil sigma model fields in homogeneous components.
Relations \eqref{bcBCi} still hold. Further, one has
\begin{subequations}
\label{xaya}
\begin{align}
\bfs{x}^a(\bfs{z})&=x^a(z)+\vartheta^\alpha \eta^+{}_{\alpha}{}^a(z)
-\hbox{$\frac{1}{2}$}\vartheta^\alpha\vartheta^\beta y^+{}_{\alpha\beta}{}^a(z),
\label{xa}
\\
\bfs{y}_a(\bfs{z})&=y_a(z)+\vartheta^\alpha \eta_{\alpha a}(z)
+\hbox{$\frac{1}{2}$}\vartheta^\alpha\vartheta^\beta x^+{}_{\alpha\beta a}(z).
\label{ya}
\end{align}
\end{subequations}
Again, the ghost number of the various component fields is given by the degree of the
superfield they appear in minus the number of $\vartheta^\alpha$ they are
multiplied by. 
The covariance properties of the component fields are intricate, but they are
completely determined by those of the superfield which they belong to.
Again, the choice of the signs is conventional.

The action and the BV variations of the Poisson--Weil sigma model
(cf. eqs. \eqref{SPW}, \eqref{dPWsuperfields}) can be written down 
explicitly in terms of the components fields. The resulting expression are
rather messy  and are collected in appendix \ref{sec:comppoisweil} for convenience.

It is interesting to study the classical version of the Poisson--Weil model and compare it
with that of the ordinary Poisson model. 
As for the classical Weil sigma model,
the classical Poisson--Weil sigma model is obtained truncating the field content 
of the full Poisson--Weil sigma model to the ghost number $0$ sector. 
The classical action of the model is then found to be 
\begin{equation}
S_{PWc}=S_{Wc}+\int_\Sigma\Big[
-\mu_i(x)B^{+i}
+\eta_aD_Ax^a+\hbox{$\frac{1}{2}$}P^{ab}(x)\eta_a\eta_b\Big],
\label{SPWcl}
\end{equation}
where the classical Weil sigma model action $S_{Wc}$ is given in \eqref{SWcl}. 
Again, as for the classical Weil sigma model, 
the symmetry variations of classical Poisson--Weil sigma model are
obtained from the BV variations of the full Poisson--Weil sigma model by retaining 
only the ghost fields of ghost number $1$,
\begin{subequations}
\label{dPWcWeil}
\begin{align}
\delta_{PWc}A^i&=\delta_{Wc}A^i, \hskip7.1cm
\label{dPWcAi}
\\
\delta_{PWc}b_i&=\delta_{Wc}b_i-u_i{}^a(x)y_a,
\label{dPWcbi}
\\
\delta_{PWc}B^{+i}&=\delta_{Wc}B^{+i},
\label{dPWcB+i}
\\
\delta_{PWc}c^i&=\delta_{Wc}c^i,
\label{dPWcci}
\\
\delta_{PWc}\psi^i&=\delta_{Wc}\psi^i,
\label{dPWcpsii}
\\
\delta_{PWc}x^a&=P^{ab}(x)y_b+u_i{}^a(x)c^i,
\label{dPWcxa}
\\
\delta_{PWc}\eta_a&=D_Ay_a+\partial_aP^{bc}(x)\eta_by_c+\partial_au_i{}^b(x)\eta_bc^i
-\partial_a\mu_i(x)\psi^i
\label{dPWcetaa}
\\
\delta_{PWc}y_a&=\hbox{$\frac{1}{2}$}\partial_aP^{bc}(x)y_by_c
+\partial_au_i{}^b(x)y_bc^i,
\label{dWcya}
\end{align}
\end{subequations}
where the classical Weil sigma model $\delta_{Wc}$ variations are given 
by \eqref{dWcWeil}.
One check that $S_{PWc}$ is invariant under the above field variations,
\begin{equation}
\delta_{PWc}S_{PWc}=0.
\end{equation}
The classical field variation operator $\delta_{PWc}$ is nilpotent but only on--shell,
\begin{equation}
\delta_{PWc}{}^2=0 \qquad\hbox{on--shell}.
\end{equation}

\vfill\eject

\begin{small}
\section{  \bf The gauge fixing of the Poisson--Weil model}
\label{sec:poissonweilfixing}
\end{small}

In this section, we carry out the gauge fixing of the Poisson--Weil sigma model.
Unlike the gauge fixing of the Weil sigma model, which is essentially unique, 
the gauge fixing of the Poisson--Weil sigma model can in principle be carried out in
several generally inequivalent ways depending on the nature of the target
space geometry. Exploring all the possibilities is out question. Below, we concentrate
on a gauge fixing prescription that leads to an interesting topological field theory.

We assume that the data defining the Poisson--Weil sigma model 
satisfy the following additional requirements.

\vskip.35cm

\begin{enumerate}

\item The manifold $M$ is endowed with a Kaehler structure.

\item The $G$--action on $M$ preserves the Kaehler structure.

\item The $G$--invariant $2$--vector $P$ is the one canonically associated with
the Kaehler structure. 

\end{enumerate}

By a Kaehler structure, we mean a pair $(J,g)$ formed by an almost complex
structure $J$ and a Riemannian metric $g$, such that $g$ is Hermitian 
with respect to $J$ and $J$ is parallel with respect to the Levi--Civita
connection of $g$. The almost complex structure $J$ 
is then automatically integrable and, thus, a complex structure.
The Kaehler form $\omega=gJ$ defines a symplectic structure and thus
a Poisson structure $P=\omega^{-1}$. Explicitly
\begin{equation}
P^{rs}=0,\qquad P^{r\bar s}=-ig^{r\bar s} \qquad \hbox{and c. c.} 
\end{equation}
The $G$--invariance of the Kaehler structure
entails the $G$--invariance of $P$. 

As explained in sect. \ref{sec:gaugedpoisson}, the consistency of the model
requires the $G$--action to be Hamiltonian with moment map $\mu$. Explicitly,
\begin{equation}
u^r{}_i=ig^{r\bar s}\partial_{\bar s}\mu_i \qquad  \hbox{and c. c.} 
\label{uham}
\end{equation}

The invariance of the Kaehler structure under the $G$--action entails
that the fundamental vector field $u$ of the $G$--action is both holomorphic and Killing. 
This leads to the relations
\begin{subequations}
\label{uholkill}
\begin{align}
&\nabla_{\bar r}u^s{}_i=0 \qquad  \hbox{and c. c.}, 
\label{uhol}
\\
&\nabla_{\bar r}u^{\bar s}{}_i+g_{\bar r t}g^{\bar s u}\nabla_uu^t{}_i=0 \qquad  \hbox{and c. c.}. 
\label{ukill}
\end{align}
\end{subequations}
Combining \eqref{uham}, \eqref{uhol}, \eqref{ukill}, one finds that $\mu$ must
satisfy the equation
\begin{equation}
\nabla_r\partial_s\mu_i=0\qquad  \hbox{and c. c.}.
\label{nablanablamu=0}
\end{equation}

Proceeding in a way analogous to that of the Weil sigma model, 
the gauge fixing is implemented by adding the
auxiliary fields of the Weil sigma model (cf. sect. \ref{sec:yangmillsfixing})
to the field content of the Poisson--Weil sigma model and by adding the auxiliary field action
$S_{W\mathrm{aux}}$ (cf. sect. \ref{SWaux})
to the Poisson--Weil sigma model action $S_{PW}$: \hphantom{xxxxxxxxxxxxxxxxxx}
\begin{equation}
S_{PW\mathrm{ext}}=S_{PW}+S_{W\mathrm{aux}}.
\end{equation}
The gauge fixed action $I_{PW}$ is obtained by restricting $S_{PW\mathrm{ext}}$ to a
suitable Lagrangian submanifold $\mathfrak{L}_{PW}$ in field space,
\begin{equation}
I_{PW}=S_{PW\mathrm{ext}}\big|_{\mathfrak{L}_{PW}}.
\label{I_PWdef}
\end{equation}
$I_{PW}$ is invariant under a BRST symmetry $s_{PW}$, which is the residual BV
symmetry left intact by the gauge fixing.

The gauge fixing requires, among other things, the choice of a metric
of $\Sigma$. In this way, as is well-known, $\Sigma$ acquires in
canonical fashion a complex structure. The tangent bundle of $\Sigma$ 
splits then in its holomorphic and antiholomorphic components
$T\Sigma=T^{(1,0)}\Sigma\oplus T^{(0,1)}\Sigma$ and similarly for the cotangent
bundle. Henceforth, 
we conveniently redefine our notation according to $\phi^{(1,0)}\rightarrow\phi_\mathrm{c}$,
$\phi^{(0,1)}\rightarrow\overline{\phi}{}_\mathrm{c}$ for a given $1$--form field 
$\phi\in\Omega^*(\Sigma)$. 

We implement the gauge fixing, by using the gauge fixing conditions
\eqref{LWdef} previously employed in the Weil sector of the model
and the further conditions 
\begin{subequations}
\label{LPWsdef}
\begin{align}
\eta_{\mathrm{c}r}&=0 \qquad \hbox{and c. c.},
\label{}
\\
\eta^+{}_\mathrm{c}{}^r&=0 \qquad \hbox{and c. c.},
\label{}
\\
x^+{}_r&=0\qquad \hbox{and c. c.},
\label{}
\\
y^{+r}&=0 \qquad \hbox{and c. c.},
\label{}
\end{align}
\end{subequations}
\cite{AKSZ,Bonechi}.
Using \eqref{OmegaPW}, it is easy to see that these define a Lagrangian
submanifold $\mathfrak{L}_{PW}$ in field space.
Note that, unlike the Weil sigma model, the condition \eqref{LPWsdef} are not derived directly
from a gauge fermion, but that does not matter as long as $\mathfrak{L}_{PW}$
is Lagrangian as required.

After a  computation, we find
\begin{align}
I_{PW}&=I_W                 
+\int_\Sigma\Big[ig_{\bar rs}(x)\overline{D}{}_{A\mathrm{c}}x^{\bar r}D_{A\mathrm{c}}x^s
\label{IPW}
\\
&~~~~~~~~~~~~
+\overline{\eta}{}^+{}_\mathrm{c}{}^r
\big(D_{\nabla \! A\mathrm{c}}y_r-\partial_r\mu_i(x)\psi_\mathrm{c}{}^i\big)
+\eta^+{}_\mathrm{c}{}^{\bar r}\big(\overline{D}_{\nabla \! A\mathrm{c}}y_{\bar r}
-\partial_{\bar r}\mu_i(x)\overline{\psi}{}_\mathrm{c}{}^i\big)
\vphantom{\int_\Sigma}\nonumber
\\
&~~~~~~~~~~~~
+\overline{\eta}{}^+{}_\mathrm{c}{}^r\eta^+{}_\mathrm{c}{}^{\bar s}
\big(-iR^{t\bar u}{}_{r\bar s}(x)y_ty_{\bar u}
+\partial_r\partial_{\bar s}\mu_i(x)C^i\big)
\vphantom{\int_\Sigma}\nonumber
\\
&~~~~~~~~~~~~
+b_ih^{ij}\mu_j(x)*1-B_ih^{ij}\big(u^r{}_j(x)y_r 
+u^{\bar r}{}_j(x)y_{\bar r}\big)*1\Big],
\vphantom{\int_\Sigma}\nonumber
\end{align}
where $I_W$ is the gauge fixed Weil sigma model action (cf. eq. \eqref{I_W})
and $D_{A\mathrm{c}}$, $\overline{D}{}_{A\mathrm{c}}$ are the holomorphic and
antiholomorphic component of the gauge covariant derivative operator $D_A$
(cf. eqs. \eqref{DAXY}, \eqref{dAfields}, \eqref{dAfields+}) and we have defined
\begin{equation}
D_{\nabla \! A\mathrm{c}}y_r=D_{A\mathrm{c}}y_r-\Gamma^s{}_{tr}(x)D_{A\mathrm{c}}x^ty_s
\qquad \hbox{and c. c.},
\label{}
\end{equation}
which is both gauge and general coordinate covariant
(see appendix \ref{sec:covariance}.). In the above expression,
wedge product of forms is understood again. 
Expression \eqref{IPW} is obtained upon eliminating the fields
\begin{equation}
\overline{\eta}{}'{}_{\mathrm{c}r}=\overline{\eta}{}_{\mathrm{c}r}
-\Gamma^u{}_{tr}(x)\overline{\eta}{}^+{}_\mathrm{c}{}^ty_u
-ig_{r\bar s}(x)\overline{D}{}_{A\mathrm{c}}x^{\bar s}
\qquad \hbox{and c. c.},
\label{}
\end{equation}
which decouple from all the other.

The Poisson--Weil sigma model BRST variations of the fields are obtained from \eqref{dWWeil},
\eqref{dWaux}, \eqref{dPWfields} upon restriction to $\mathfrak{L}_{PW}$. They
read 
\begin{subequations}
\label{poissonweilBRST}
\begin{align}
s_{PW}A^i&=s_WA^i,
\label{}
\\
s_{PW}\psi^i&=s_W\psi^i,
\label{}
\\
s_{PW}b_i&=s_Wb_i-u^r{}_i(x)y_r-u^{\bar r}{}_i(x)y_{\bar r},
\label{}
\\
s_{PW}B_i&=s_WB_i-\mu_i(x)
\label{}
\\
s_{PW}c^i&=s_Wc^i,
\label{}
\\
s_{PW}C^i&=s_WC^i,
\label{}
\\
s_{PW}\tilde c^i&=s_W\tilde c^i,
\label{}
\\
s_{PW}\gamma^i&=s_W\gamma^i,
\label{}
\\
s_{PW}\tilde C^i&=s_W\tilde C^i,
\label{}
\\
s_{PW}\Gamma^i&=s_W\Gamma^i,
\label{}
\\
s_{PW}x^r&=-ig^{r\bar s}(x)y_{\bar s}+u^r{}_i(x)c^i \qquad \hbox{and c.c.},
\label{}
\\
s_{PW}y_r&=\Gamma^s{}_{tr}(x)\big(-ig^{t\bar u}(x)y_{\bar u}+u^t{}_i(x)c^i\big)y_s
\label{}
\\
&~~~~~~~~~~~~~~~~~
+\nabla_ru^s{}_i(x)y_sc^i+\partial_r\mu_i(x)C^i
\qquad \hbox{and c.c.},
\nonumber
\\
s_{PW}\overline{\eta}{}^+{}_\mathrm{c}{}^r&=
-\Gamma^r{}_{ts}(x)\big(-ig^{t\bar u}(x)y_{\bar u}+u^t{}_i(x)c^i\big)
\overline{\eta}{}^+{}_\mathrm{c}{}^s
\label{}
\\
&~~~~~~~~~~~~~~~~~
+\overline{D}{}_{A\mathrm{c}}x^r+\nabla_su^r{}_i(x)\overline{\eta}{}^+{}_\mathrm{c}{}^sc^i
\qquad \hbox{and c.c.},
\nonumber
\end{align}
\end{subequations}
where the Weil sigma model $s_W$ BRST variations are given by \eqref{weilBRST}.
One can verify directly that $I_{PW}$ is BRST invariant
\begin{equation}
s_{PW}I_{PW}=0.
\label{sPWIPW=0}
\end{equation}
Further, one has \hphantom{xxxxxxxxxxxxxxxxxxxxxxx}
\begin{equation}
s_{PW}{}^2=0\qquad \hbox{on shell}.
\label{sPW2=0}
\end{equation}
Unlike the Weil sigma model, the Poisson--Weil BRST variation operator is nilpotent
only on--shell. 

It us easy to see that the field theory we have obtained by gauge fixing 
is topological. One defines a ghost number $-1$ gauge fermion $\Psi_{PW}$ by
\begin{equation}
\Psi_{PW}=\Psi_W+\int_\Sigma\Big[
\hbox{$\frac{1}{2}$}ig_{r\bar s}(x)\overline{\eta}{}^+{}_\mathrm{c}{}^rD_{A\mathrm{c}}x^{\bar s}
-\hbox{$\frac{1}{2}$}ig_{\bar r s}(x)\eta^+{}_\mathrm{c}{}^{\bar r}\overline{D}{}_{A\mathrm{c}}x^s
\Big],
\label{PsiPW}
\end{equation}
where $\Psi_W$ is the gauge fermion of the Weil sigma model given by \eqref{PsiW}.
Using \eqref{poissonweilBRST}, it can be verified that 
\begin{equation}
I_{PW}=S_{PW\mathrm{top}}+s_{PW}\Psi_{PW} \qquad \hbox{on shell},
\label{IPW=SPWtop+sPWPsi}
\end{equation}
where the Poisson--Weil topological action $S_{W\mathrm{top}}$ is given by
\begin{equation}
S_{PW\mathrm{top}}=S_{W\mathrm{top}}                 
+\int_\Sigma x^*{}_A\omega,
\label{SPWtop}
\end{equation}
with the Weil topological action $S_{W\mathrm{top}}$ given by \eqref{SWtop}.
The globally defined $2$--form
$x^*{}_A\omega$ is the gauge covariant pull-back of $\omega$ 
\begin{equation}
x^*{}_A\omega=\hbox{$\frac{1}{2}$}\omega_{ab}(x)D_Ax^aD_Ax^b
=x^*\omega+d(\mu_i(x)A^i)-\mu_i(x)F^i.
\label{}
\end{equation}
The above expression is obtained by using, among other things, the remarkable gauge
covariant Kaehler identity
\begin{equation}
ig_{\bar rs}(x)\overline{D}{}_{A\mathrm{c}}x^{\bar r}D_{A\mathrm{c}}x^s
-ig_{r\bar s}(x)\overline{D}{}_{A\mathrm{c}}x^rD_{A\mathrm{c}}x^{\bar s}=x^*{}_A\omega.
\label{}
\end{equation}
This calculation shows the topological nature of the theory. 
All dependence on the metric of $\Sigma$ and the background connection $A_0$ 
is again buried inside the gauge fermion $\Psi_{PW}$. The topological quantum field 
correlators, therefore, are going to be independent from these data. 

The topological field theory which we are dealing with has been studied  
by Baptista in a series of papers \cite{Baptista1,Baptista2,Baptista3}.
It describes the moduli space of solutions of the so called vortex equations
\cite{C-G-S,C-G-M-S,G-S,MiR1,MiR2,MiR3}. 
Strictly speaking, the sigma model Lagrangian obtained above differs from
Baptista's. However, this may be simply a gauge fixing artifact. The fact that
our sigma model and Baptista's have the same field content and localize on the 
same space of field configurations, as we show momentarily, 
indicates that they are the same topological field theory. 
As well known, topological field theories have BRST exact Lagrangians and 
this makes them invariant as field theories under large classes of 
deformations. So, it is not surprising that the same topological field 
theory may have several Lagrangian realizations. What characterizes a
topological field theory is its field content and the space of field
configurations on which the field theory localizes. Indeed, the path integral 
of the field theory is just a complicated way of writing a functional Dirac 
delta with support on such configurations. Thus, one expects two such theories sharing  
the same set of fields and localizing on the same field configurations to 
be equivalent. However, a complete proof of this statement would require
an in depth analysis of the BRST cohomologies of the two theories and a proof of
their equivalence, a task which is beyond the scope of the present paper. 

Let us now show that the topological field theory we have obtained describes 
the moduli space of solutions of the vortex equations, as claimed in 
the previous paragraph. To begin with, we note that the geometrical data
of the gauge fixed Poisson--Weil sigma model are precisely the same as those 
of the vortex equations: a principal $G$--bundle $Q$ over a
Riemann surface $\Sigma$ and a Kaehler manifold $M$ with a Hamiltonian
effective action preserving the Kaehler structure. 
The field configurations, on which our topological field theory localizes,
are the BRST invariant purely bosonic on--shell configurations. They 
are easily obtained from the expression \eqref{poissonweilBRST} 
of the BRST variations obtained above. Setting all the fermionic fields to zero in 
\eqref{poissonweilBRST} 
and imposing  that the resulting expressions vanish on shell leads to the equations
\hphantom{xxxxxxxxxxxxxxxxxxxx}
\begin{align}
&F^i+h^{ij}\mu_j(x)*1=0,
\label{vortexF}
\\
&\overline{D}{}_{A\mathrm{c}}x^r=0,
\label{vortexx}
\end{align}
which are precisely the vortex equations. 

The vortex configurations are extrema of the energy functional
\begin{equation}
\mathcal{E}
=\int_\Sigma\Big[\hbox{$\frac{1}{2}$}h_{ij}F^i*F^j+
\hbox{$\frac{1}{2}$}g_{ab}(x)D_Ax^a*D_Ax^b+\hbox{$\frac{1}{2}$}h^{ij}\mu_i\mu_j(x)*1\Big],
\end{equation}
first written down in \cite{C-G-S,MiR3}. 
However, they are not generic extrema. They are instanton like energy
minimizing configurations. Indeed, by means of Bogomolny type manipulations,
one can show that $\mathcal{E}$ can be written as
\begin{align}
\mathcal{E}&=\eta_{E_M}
+\int_\Sigma\Big[-2ig_{r\bar s}(x)\overline{D}{}_{A\mathrm{c}}x^rD_{A\mathrm{c}}x^{\bar s}
\label{Bogom}
\\
&\hskip 2cm
+\hbox{$\frac{1}{2}$}h_{ij}
\big(F^i+h^{ik}\mu_k(x)*1\big)*\big(F^j+h^{jl}\mu_l(x)*1\big)
\Big],
\nonumber
\end{align}
where $\eta_{E_M}$ is given by 
\begin{equation}
\eta_{E_M}=-\int_\Sigma\Big[x^*\omega+d(\mu_i(x)A^i)\Big].
\end{equation}
$\eta_{E_M}$ depends only on the homotopy class of
$x$ and is independent from $A$. It is thus a topological invariant characterizing
the bundle $E_M$. The remaining term in the right hand side of \eqref{Bogom}
is positive definite and vanishes precisely, when the vortex equations
\eqref{vortexF}, \eqref{vortexx} are satisfied. Thus, the energy $\mathcal{E}$ 
is minimized by the vortex configurations and the minimum equals the 
topological invariant $\eta_{E_M}$. 
In this sense, vortex configurations are akin to instantons. 

Our gauged topological sigma model can be viewed as a topological field
theoretic completion of a purely bosonic theory with action $\mathcal{E}$.
Indeed, the ghost number $0$ sector $I_{PW}{}^0$ of the action $I_{PW}$ after
algebraically eliminating the auxiliary field $b$ is given by \hphantom{xxxxxxxxxxxxxxx}
\begin{equation}
I_{PW}{}^0=-\hbox{$\frac{1}{2}$}\eta_{E_M}-\hbox{$\frac{1}{2}$}\mathcal{E}.
\end{equation}

The topological sigma model, which we have obtained, is in fact the gauged
version  of Witten's $A$--model originally worked out in \cite{Witten1,Witten2}.
In the case where the group $G$ is trivial, 
the action $I_{PW}$ reduces indeed 
to the well-known action  of the $A$--model \cite{Bonechi}. 

The $A$--model is known to be related to the quantum cohomology of the target
manifold $M$: its correlators compute the Gromov--Witten invariants.  
The importance of the vortex equation moduli space stems from the 
realization that it enters the definition of the Hamiltonian Gromov--Witten
invariants \cite{C-G-M-S}.

\vfill\eject

\begin{small}
\section{\bf Outlook}
\label{sec:outlook}
\end{small}

The constructions expounded in this paper are likely to be extendable in 
several directions. 

We have formulated the Weil sigma model for a principal $G$--bundle $Q$ over
$\Sigma$ with $G$ a Lie group. One possibility would be to generalize the model 
to the case where $G$ is a Poisson--Lie group. One expects the Lie bialgebra 
structure of $\mathfrak{g}$ to play a basic role in this case. The Weil sigma model 
described in the paper would be the special case where $G$ has the trivial Poisson 
structure. 

As a further step, one may try to couple the generalized sigma Weil model so obtained
to the Poisson sigma model with target space $M$ carrying a Hamiltonian Poisson
action of the Poisson--Lie group $G$. This would yield a generalized 
Poisson--Weil sigma model and would be the gauging of the Poisson sigma model by the  
Poisson--Lie $G$--symmetry 
\footnote{$\vphantom{\bigg[}$ This possibility was suggested to us by F. Bonechi.}.

The basic and equivariant Poisson--Lichnerowicz cohomology of $M$ have been defined
and studied by Ginzburg \cite{Ginzburg} also for this more general setting.  
Note that the moment map $\mu$ would be $G^\vee$--valued rather than
$\mathfrak{g}^\vee$--valued in this case, where $G^\vee$ is dual Poisson--Lie
dual group of $G$. 
The BV cohomology of the generalized Poisson--Weil sigma 
model should again be related to this more general cohomology.

The Poisson sigma model with a Poisson--Lie target space $G$ has been
studied in \cite{Calvo1, Calvo2, Bonechi2}. One may explore
the relation of these models with the one resulting from the 
constructions just outlined.

It remains to be seen whether the generalized models are going to yield
interesting topological field theories upon gauge fixing. All this is left to
future work.

\vfill\eject 

\appendix

\begin{small}
\section{\bf The Weil model in components}
\label{sec:compweil}
\end{small}

In this appendix, we collect the explicit expressions of the action and the BV symmetry
transformations of the Weil sigma model in terms of the component fields. 

The expansion of the Weil sigma model superfields in components 
is given in \eqref{bcBCi}.
The ghost numbers of the components are given by the following table:
\begin{equation}\label{weilghno}
\begin{matrix}
&\gh b_i=0,\hfill\hfill&\gh A^+{}_i=-1,\hfill\hfill&\gh c^+{}_i=-2,\hfill\hfill
\\
&\gh c^i=+1,\hfill\hfill&\gh A^i=0,\hfill\hfill&\gh b^{+i}=-1,\hfill\hfill
\\
&\gh B_i=-1,\hfill\hfill&\gh \psi^+{}_i=-2,\hfill\hfill&\gh C^+{}_i=-3,\hfill\hfill
\\
&\gh C^i=+2,\hfill\hfill&\gh\psi^i=+1,\hfill\hfill&\gh B^{+i}=0.\hfill\hfill
\end{matrix}
\end{equation}

The Weil sigma model action $S_W$ (cf. eq. \eqref{SW}) in components reads
\begin{align}
S_W&=\int_\Sigma\Big[-b_i\big(F_A{}^i+B^{+i}-f^i{}_{jk}b^{+j}c^k\big)
+A^+{}_i\big(D_Ac^i-\psi^i)
\label{SWcomp}
\\
&~~~~~~~~~~
+B_i(D_A\psi^i-f^i{}_{jk}c^jB^{+k}-f^i{}_{jk}b^{+j}C^k\big)
-\psi^+{}_i\big(D_AC^i+f^i{}_{jk}c^j\psi^k\big)
\vphantom{\int_\Sigma}\nonumber
\\
&~~~~~~~~~~
+c^+{}_i\big(C^i-\hbox{$\frac{1}{2}$}f^i{}_{jk}c^jc^k\big)+C^+{}_if^i{}_{jk}c^jC^k\Big],
\vphantom{\int_\Sigma}
\nonumber
\end{align}
where \hphantom{xxxxxxxxxxxxxxxxxxxxxxxxxxxxxx}
\begin{equation}
F_A{}^i=dA^i+\hbox{$\frac{1}{2}$}f^i{}_{jk}A^jA^k
\label{FAi}
\end{equation}
is the curvature of the connection $A$ and 
\begin{subequations}
\label{DAXY}
\begin{align}
D_AX^i&=dX^i+f^i{}_{jk}A^jX^k,
\label{DAXi}
\\
D_AY_i&=dY_i-f^k{}_{ji}A^jY_k
\label{DAYi}
\end{align}
\end{subequations}
are the gauge covariant derivatives of $X\in
\Omega^*(\Sigma,\mathrm{Ad}\,Q)$ and $Y\in\Omega^*(\Sigma,\mathrm{Ad}^\vee Q)$, 
respectively. Above, the various fields are local forms on $\Sigma$ obtained
by the corresponding components of the basic superfields by the formal 
replacement $\vartheta^\alpha\rightarrow dz^\alpha$. Wedge multiplication of
forms is understood.

The Weil sigma model BV variations (cf. eq. \eqref{dWsuperfields}) of the
components are explicitly given by 
\begin{subequations}
\label{dWWeil}
\begin{align}
\delta_Wc^i&=C^i-\hbox{$\frac{1}{2}$}f^i{}_{jk}c^jc^k,
\label{dWci}
\\
\delta_WA^i&=\psi^i-D_Ac^i,
\label{dWAi}
\\
\delta_Wb^{+i}&=B^{+i}+F_A{}^i-f^i{}_{jk}c^jb^{+k},
\label{dWb+i}
\\
\delta_Wb_i&=f^k{}_{ji}c^jb_k+f^k{}_{ji}C^jB_k,
\label{dWbi}
\\
\delta_WA^+{}_i&=D_Ab_i+f^k{}_{ji}c^jA^+{}_k+f^k{}_{ji}C^j\psi^+{}_k-f^k{}_{ji}\psi^jB_k,
\label{dWA+i}
\\
\delta_Wc^+{}_i&=D_AA^+{}_i-f^k{}_{ji}b^{+j}b_k+f^k{}_{ji}c^jc^+{}_k
\label{dWc+i}
\\
&~~~~~~~~-f^k{}_{ji}\psi^j\psi^+{}_k
-f^k{}_{ji}B^{+j}B_k+f^k{}_{ji}C^jC^+{}_k,
\nonumber
\\
\delta_WC^i&=-f^i{}_{jk}c^jC^k,
\label{dWCi}
\\
\delta_W\psi^i&=-D_AC^i-f^i{}_{jk}c^j\psi^k,
\label{dWpsii}
\\
\delta_WB^{+i}&=D_A\psi^i-f^i{}_{jk}c^jB^{+k}-f^i{}_{jk}b^{+j}C^k,
\label{dWB+i}
\\
\delta_WB_i&=-b_i+f^k{}_{ji}c^jB_k
\label{dWBi}
\\
\delta_W\psi^+{}_i&=-A^+{}_i+D_AB_i+f^k{}_{ji}c^j\psi^+{}_k,
\label{dWpsi+i}
\\
\delta_WC^+{}_i&=D_A\psi^+{}_i+f^k{}_{ji}c^jC^+{}_k-f^k{}_{ji}b^{+j}B_k-c^+{}_i,
\label{dWC+i}
\end{align}
\end{subequations}
as one can check by a simple computation.

\vfill\eject

\begin{small}
\section{\bf The Poisson--Weil model in components}
\label{sec:comppoisweil}
\end{small}

In this appendix, we collect the explicit expressions of the action and the BV symmetry
transformations of the Poisson--Weil sigma model in terms of the component fields. 

The expansion of the Poisson--Weil sigma model superfields in terms of components
is given in \eqref{bcBCi}, \eqref{xaya}. The ghost numbers of the components
are given by table \eqref{weilghno} and by the following one:
\begin{equation}\label{poisweilghno}
\begin{matrix}
&\gh x^a=0,\hfill\hfill&\gh \eta^{+a}=-1,\hfill\hfill&\gh  y^{+a}=-2,\hfill\hfill
\\
&\gh y_a=+1,\hfill\hfill&\gh \eta_a=0,\hfill\hfill&\gh x^+{}_a=-1.\hfill\hfill
\end{matrix}
\end{equation}

The Poisson--Weil sigma model action $S_{PW}$ (cf. eq. \eqref{SPW}) in
components is given by
\begin{align}
S_{PW}&=S_W+\int_\Sigma\Big[
\eta_aD_Ax^a+\hbox{$\frac{1}{2}$}P^{ab}(x)\eta_a\eta_b
\\
&~~~~~~~~~~~~~~~~~
+\eta^{+a}\big(D_Ay_a
+\partial_aP^{bc}(x)\eta_by_c+\partial_au_i{}^b(x)\eta_bc^i
-\partial_a\mu_i(x)\psi^i\big)
\vphantom{\int_\Sigma}\nonumber
\\
&~~~~~~~~~~~~~~~~~
+\hbox{$\frac{1}{2}$}\eta^{+a}\eta^{+b}\big(\hbox{$\frac{1}{2}$}\partial_a\partial_bP^{cd}(x)y_cy_d
+\partial_a\partial_bu_i{}^c(x)y_cc^i+\partial_a\partial_b\mu_i(x)C^i\big)
\vphantom{\int_\Sigma}
\nonumber
\\
&~~~~~~~~~~~~~~~~~
-u_i{}^a(x)y_ab^{+i}-\mu_i(x)B^{+i}+x^+{}_a\big(P^{ab}(x)y_b+u_i{}^a(x)c^i\big)
\vphantom{\int_\Sigma}
\nonumber
\\
&~~~~~~~~~~~~~~~~~
-y^{+a}\big(\hbox{$\frac{1}{2}$}\partial_aP^{bc}(x)y_by_c
+\partial_au_i{}^b(x)y_bc^i+\partial_a\mu_i(x)C^i\big)
\vphantom{\int_\Sigma}\Big],
\nonumber
\end{align}
where $S_W$ is given by \eqref{SWcomp} and 
\begin{subequations}
\label{dAfields}
\begin{align}
D_Ax^a&=dx^a-u_i{}^a(x)A^i,
\label{dAxa}
\\
D_Ay_a&=dy_a+\partial_au_i{}^b(x)A^iy_b
\label{dAya}
\end{align}
\end{subequations}
are the gauge covariant derivatives of $x$ and $y$, respectively.
Recall that the various fields are local forms on $\Sigma$ 
obtained from the corresponding components of the superfields by the formal replacement 
$\vartheta^\alpha\rightarrow dz^\alpha$ and that wedge multiplication of
forms is understood throughout. The main properties of the gauge covariant
derivatives are collected in appendix \ref{sec:covariance}.

The Poisson--Weil sigma model BV variations (cf. eq. \eqref{dPWsuperfields}) of the
component fields are given by 
\begin{subequations}
\label{dPWfields}
\begin{align}
\delta_{PW}c^i&=\delta_Wc^i,
\label{dPWci}
\\
\delta_{PW}A^i&=\delta_WA^i,
\label{dPWAi}
\\
\delta_{PW}b^{+i}&=\delta_Wb^{+i},
\label{dPWb+i}
\\
\delta_{PW}b_i&=\delta_Wb_i-u_i{}^a(x)y_a,
\label{dPWbi}
\\
\delta_{PW}A^+{}_i&=\delta_WA^+{}_i
-\partial_au_i{}^b(x)\eta^{+a}y_b-u_i{}^a(x)\eta_a,
\label{dPWA+i}
\\
\delta_{PW}c^+{}_i&=\delta_Wc^+{}_i
-\partial_au_i{}^b(x)\eta^{+a}\eta_b-u_i{}^a(x)x^+{}_a
\label{dPWc+i}
\\
&\hskip3.0cm
-\hbox{$\frac{1}{2}$}\partial_a\partial_bu_i{}^c(x)\eta^{+a}\eta^{+b}y_c
+\partial_au_i{}^b(x)y^{+a}y_b,
\nonumber
\\
\delta_{PW}C^i&=\delta_WC^i,
\label{dPWCi}
\\
\delta_{PW}\psi^i&=\delta_W\psi^i,
\label{dPWpsii}
\\
\delta_{PW}B^{+i}&=\delta_WB^{+i},
\label{dPWB+i}
\\
\delta_{PW}B_i&=\delta_WB_i-\mu_i(x),
\label{dPWBi}
\\
\delta_{PW}\psi^+{}_i&=\delta_W\psi^+{}_i-\partial_a\mu_i(x)\eta^{+a},
\label{dPWpsi+i}
\\
\delta_{PW}C^+{}_i&=\delta_WC^+{}_i
-\hbox{$\frac{1}{2}$}\partial_a\partial_b\mu_i(x)\eta^{+a}\eta^{+b}
+\partial_a\mu_i(x)y^{+a},
\label{dPWC+i}
\\
\delta_{PW}x^a&=P^{ab}(x)y_b+u_i{}^a(x)c^i,
\label{dPWxa}
\\
\delta_{PW}\eta^{+a}&=D_Ax^a+\partial_cP^{ab}(x)\eta^{+c}y_b
+\partial_bu_i{}^a(x)\eta^{+b}c^i+P^{ab}(x)\eta_b,
\label{dPWeta+a}
\\
\delta_{PW}y^{+a}&=-D_A\eta^{+a}                
-\partial_cP^{ab}(x)\eta^{+c}\eta_b
\label{dPWy+a}
\\
&~~~-\hbox{$\frac{1}{2}$}\partial_c\partial_dP^{ab}(x)\eta^{+c}\eta^{+d}y_b
-\hbox{$\frac{1}{2}$}\partial_b\partial_cu_i{}^a(x)\eta^{+b}\eta^{+c}c^i
\nonumber
\\
\hskip1.7cm
&~~~-P^{ab}(x)x^+{}_b+\partial_cP^{ab}(x)y^{+c}y_b
+\partial_bu_i{}^a(x)y^{+b}c^i+u_i{}^a(x)b^{+i},
\nonumber
\\
\delta_{PW}y_a&=\hbox{$\frac{1}{2}$}\partial_aP^{bc}(x)y_by_c
+\partial_au_i{}^b(x)y_bc^i+\partial_a\mu_i(x)C^i,
\label{dWya}
\\
\delta_{PW}\eta_a&=D_Ay_a            
+\hbox{$\frac{1}{2}$}\partial_a\partial_dP^{bc}(x)\eta^{+d}y_by_c
+\partial_aP^{bc}(x)\eta_by_c
\label{dPWetaa}
\\
\hskip1.7cm
&~~~
+\partial_a\partial_cu_i{}^b(x)\eta^{+c}y_bc^i+\partial_au_i{}^b(x)\eta_bc^i
\nonumber
\\
\hskip1.7cm
&~~~-\partial_a\mu_i(x)\psi^i+\partial_a\partial_b\mu_i(x)\eta^{+b}C^i,
\nonumber
\\
\delta_{PW}x^+{}_a&=D_A\eta_a                  
+\hbox{$\frac{1}{2}$}\partial_aP^{bc}(x)\eta_b\eta_c
-\partial_a\partial_cu_i{}^b(x)\eta^{+c}y_bA^i
\label{dPWx+a}
\end{align}
\vfill\eject
\begin{align}
\hskip1.7cm
&~~~+\partial_a\partial_cu_i{}^b(x)\eta^{+c}\eta_bc^i
+\partial_a\partial_dP^{bc}(x)\eta^{+d}\eta_by_c
\nonumber
\\
\hskip1.7cm
&~~~
+\hbox{$\frac{1}{4}$}\partial_a\partial_d\partial_eP^{bc}(x)\eta^{+d}\eta^{+e}y_by_c
+\hbox{$\frac{1}{2}$}\partial_a\partial_c\partial_du_i{}^b(x)\eta^{+c}\eta^{+d}y_bc^i
\nonumber
\\
\hskip1.7cm
&~~~
+\partial_aP^{bc}(x)x^+{}_by_c
+\partial_au_i{}^b(x)x^+{}_bc^i
-\hbox{$\frac{1}{2}$}\partial_a\partial_dP^{bc}(x)y^{+d}y_by_c
\nonumber
\\
\hskip1.7cm
&~~~
-\partial_a\partial_cu_i{}^b(x)y^{+c}y_bc^i-\partial_au_i{}^b(x)y_bb^{+i}
-\partial_a\partial_b\mu_i(x)\eta^{+b}\psi^i
\nonumber
\\
\hskip1.7cm
&~~~
+\hbox{$\frac{1}{2}$}\partial_a\partial_b\partial_c\mu_i(x)\eta^{+b}\eta^{+c}C^i
-\partial_a\partial_b\mu_i(x)y^{+b}C^i-\partial_a\mu_i(x)B^{+i},
\nonumber
\end{align}
\end{subequations}
where the Weil sigma model $\delta_W$ variations are given by \eqref{dWWeil} and 
\begin{subequations}
\label{dAfields+}
\begin{align}
D_A\eta^{+a}&=d\eta^{+a}-\partial_bu_i{}^a(x)A^i\eta^{+b},
\label{dAeta+}
\\
D_A\eta_a&=d\eta_a+\partial_au_i{}^b(x)A^i\eta_b
\label{dAeta}
\end{align}
\end{subequations}
are the gauge covariant derivatives of $\eta^+$ and $\eta$, respectively
\footnote{$\vphantom{\bigg[}$ Here, we are abusing our terminology. Strictly
speaking, $D_A\eta$, as defined above, is gauge covariant only when $y$
vanishes. See again appendix \ref{sec:covariance}.}.

\vfill\eject

\begin{small}
\section{\bf $G$ and general covariance}
\label{sec:covariance}
\end{small}

The covariance of the fields of the sigma models studied in the main body
of the paper is rather intricate. The embedding field is not simply a map
$x:\Sigma \rightarrow M$, as in the ordinary ungauged sigma models, but a 
section of the bundle $E_M=Q\times_G M$, whose definition combines in a non 
trivial manner the principal $G$--bundle $Q$ on $\Sigma$ and the manifold $M$ 
with $G$--action. The other fields are sections of bundles which are (related
to) pull-backs by $x$ of bundles on $E_M$. The construction of suitable 
gauge and general covariant derivatives of the fields is thus a subtle matter.
For the sake of concreteness, 
it may be useful to write down the covariance of these fields and their
covariant derivatives in terms of the cocycles representing the bundles, 
which they are sections of. This is done in the present appendix.
More material on this topic can be found for instance in 
\cite{Baptista1,Baptista2,Baptista3,MiR3}.

Let $G$ be a connected Lie group.
Let the manifold $M$ carry a left $G$--action. 
The fundamental vector field $u$ of the $G$--action is defined by the relation
\begin{equation}
g^a(m)=m^a-\xi^iu_i{}^a(m)+O(\xi^2),
\end{equation}
for $g=\exp(\xi)\in G$ with $\xi\in\mathfrak{g}$.
$u$ is $G$--equivariant, i. e. for $g\in G$,  
\begin{equation}
\partial_bg^{-1a}\circ gu_i{}^b\circ g=\big(\mathrm{Ad}g\big){}^j{}_iu_j{}^a.
\end{equation}

Let $Q$ be a principal $G$--bundle on the closed surface $\Sigma$. 
Let $\{g_{AB}(z)\}$ be a $G$--valued $1$--cocycle representing $Q$. 
Here, $A,~B,~C$ ... are local trivialization indices.
The $1$-cocycle condition
\begin{equation}
g_{AB}(z)g_{BC}(z)=g_{AC}(z),
\end{equation}
when defined, holds. 

Let $E_M$ be the fiber bundle on $\Sigma$ represented by the non linear cocycle
$\{g_{AB}{}^a(z,m_B)\}$ obtained from the $G$--valued $1$--cocycle $\{g_{AB}(z)\}$
representing $Q$ via the $G$--action on $M$. 
A section $x\in \Gamma(\Sigma,E_M)$ is given locally as a collection
of maps $\{x_A(z)\}$ into $M$ matching as \hphantom{xxxxxxxxxxxxxxxxxxxxxx}
\begin{equation}
x_A{}^a(z)=g_{AB}{}^a(z,x_B(z)).
\end{equation}

Let $x\in \Gamma(\Sigma,E_M)$. Let $x^*\mathrm{Vert}\,TE_M$ be the vector
bundle on $\Sigma$ represented by the $1$--cocycle 
$\{C_{AB}{}^a{}_b(z)\}$, where $C_{AB}{}^a{}_b(z)=\partial_bg_{AB}{}^a(z,x_B(z))$.
A section $v\in \Omega^0(\Sigma, x^*\mathrm{Vert}\,TE_M)$ is given locally as a collection
of 
$TM$--valued functions $\{v_A{}^a(z)\}$ matching as
\begin{equation}
v_A{}^a(z)=\partial_bg_{AB}{}^a(z,x_B(z))v_B{}^b(z).
\end{equation}
In similar fashion,  
let $x^*\mathrm{Vert}^*TE_M$ be the vector
bundle on $\Sigma$ represented by the $1$--cocycle 
$\{C^*{}_{AB}{}_a{}^b(z)\}$, where $C^*{}_{AB}{}_a{}^b(z)=\partial_ag_{BA}{}^b(z,x_A(z))$.
A section $s\in \Omega^0(\Sigma, x^*\mathrm{Vert}^*TE_M)$ is given locally as a collection
of 
$T^*M$--valued functions $\{s_A{}_a(z)\}$ matching as
\begin{equation}
s_A{}_a(z)=\partial_ag_{BA}{}^b(z,x_A(z))s_B{}_b(z).
\end{equation}

We want to construct gauge covariant derivatives for sections of the 
bundles $E_M$, $x^*\mathrm{Vert}\,TE_M$, $x^*\mathrm{Vert}^*TE_M$. 
To this end, one needs a connection of $Q$. Re\-call that 
a connection $A$ of $Q$ is given locally as a collection
of $\mathfrak{g}$--valued $1$--forms $\{A_A{}^i(z)\}$ matching as
\begin{equation}
A_A{}^i(z)=\big(\mathrm{Ad}g_{AB}(z)\big){}^i{}_jA_B{}^j(z)+\big(g_{AB}(z)dg_{AB}(z){}^{-1}\big){}^i,
\label{Amatch}
\end{equation}
where here and below $d$ denote the de Rham differential of $\Sigma$.

The following relation 
\begin{equation}
dg_{AB}{}^a(z,m_B)=\big(g_{AB}(z)dg_{AB}(z){}^{-1}\big){}^iu_i{}^a(g_{AB}(z,m_B))
\label{uncov}
\end{equation}
plays a basic role in the following analysis of covariance.
 
For $x\in \Gamma(\Sigma,E_M)$, define
\begin{equation}
D_Ax^a=dx^a-u_i{}^a(x)A^i.
\end{equation}
Using \eqref{Amatch}, \eqref{uncov}, one finds that 
\begin{equation}
(D_Ax)_A{}^a(z)=\partial_bg_{AB}{}^a(z,x_B(z))(D_Ax)_B{}^b(z).
\end{equation}
This shows that $D_Ax\in \Omega^1(\Sigma, x^*\mathrm{Vert}\,TE_M)$.
In this sense, $D_Ax$ is the gauge covariant derivative of $x$. 

Let $x\in \Gamma(\Sigma,E_M)$. For $v\in \Omega^0(\Sigma,x^*\mathrm{Vert}\,TE_M)$, 
define
\begin{equation}
D_Av^a=dv^a-\partial_bu_i{}^a(x)A^iv^b.
\label{DAv}
\end{equation}
Then, using \eqref{Amatch}, \eqref{uncov}, one finds 
\begin{align}
(D_Av)_A{}^a(z)&=\partial_bg_{AB}{}^a(z,x_B(z))(D_Av)_B{}^b(z)
\label{DAvmatch}
\\
&\hskip 3.5cm+\partial_b\partial_cg_{AB}{}^a(z,x_B(z))(D_Ax)_B{}^b(z)v_B{}^c(z).
\nonumber
\end{align}
Similarly, for $s\in \Omega^0(\Sigma, x^*\mathrm{Vert}^*TE_M)$, define
\begin{equation}
D_As_a=ds_a+\partial_au_i{}^b(x)A^is_b.
\end{equation}
Then, using \eqref{Amatch}, \eqref{uncov} again, one obtains 
\begin{align}
(D_As)_A{}_a(z)&=\partial_ag_{BA}{}^b(z,x_A(z))(D_As)_B{}_b(z)
\label{DAsmatch}
\\
&\hskip 3.5cm+\partial_a\partial_bg_{BA}{}^c(z,x_A(z))(D_Ax)_A{}^b(z)s_B{}_c(z).
\nonumber
\end{align}
Note that $D_Av\not \in \Omega^1(\Sigma, x^*\mathrm{Vert}\,TE_M)$
because of the second term in the right hand side of \eqref{DAvmatch}.
However, notice that this term would be absent if $M$ were a linear space
and the $G$--action on $M$ were linear, that is if $E_M$ were a vector bundle.
For this reason, with an abuse of language, we call $D_Av$ the gauge covariant derivative of $v$.
Similarly, $D_As\not \in \Omega^1(\Sigma, x^*\mathrm{Vert}^*TE_M)$
because of the second term in the right hand side of \eqref{DAsmatch}.
Again, with an abuse of language, we call $D_As$ the gauge covariant derivative of $s$. 

One can correct the lack of full covariance found above by using a
$G$--invariant connection of $M$. Recall that 
a connection $\Gamma$ of $TM$ is said $G$--invariant, if, for any $g\in G$
\begin{equation}
\Gamma^a{}_{bc}=\Gamma^d{}_{ef}\circ g\partial_dg^{-1a}\circ g\partial_bg^e\partial_cg^f
+ \partial_dg^{-1a}\circ g\partial_b\partial_cg^d.
\end{equation}
The Levi--Civita connection associated to a $G$--invariant Riemannian metric
is $G$--invariant.

For a section $v\in \Omega^0(\Sigma, x^*\mathrm{Vert}\,TE_M)$, we define 
\begin{equation}
D_{\nabla A}v^a=D_Av^a+\Gamma^a{}_{bc}(x)D_Ax^bv^c.
\label{DnablaAv}
\end{equation}
Then, under a change of local trivialization
\begin{equation}
(D_{\nabla A}v)_A{}^a(z)=\partial_bg_{AB}{}^a(z,x_B(z))(D_{\nabla A}v)_B{}^b(z).
\label{}
\end{equation}
Thus, $D_{\nabla A}v\in \Omega^1(\Sigma, x^*\mathrm{Vert}\,TE_M)$
and $D_{\nabla A}v$ is a genuine covariant derivative. 
Similarly, for a section $s\in \Omega^0(\Sigma, x^*\mathrm{Vert}^*TE_M)$, we define 
\begin{equation}
D_{\nabla A}s_a=D_As_a-\Gamma^c{}_{ba}(x)D_Ax^bs_c.
\label{DnablaAs}
\end{equation}
Then, under a change of local trivialization
\begin{equation}
(D_{\nabla A}s)_A{}_a(z)=\partial_ag_{BA}{}^b(z,x_A(z))(D_{\nabla A}s)_B{}_b(z).
\label{}
\end{equation}
Thus, $D_{\nabla A}s\in \Omega^1(\Sigma, x^*\mathrm{Vert}^*TE_M)$
and $D_{\nabla A}s$ is a genuine covariant derivative.

\vfill\eject


\begin{thebibliography}{99}

\bibitem{Marsden:1}
J.~Marsden and A.~Weinstein, 
``Reduction of symplectic manifolds with symmetry'', 
Rep. Math. Phys. {\bf 5} (1974) 121. 

\bibitem{Witten3}
E.~Witten,
``Phases of N = 2 theories in two dimensions'',
Nucl.\ Phys.\  B {\bf 403} (1993) 159
[arXiv:hep-th/9301042].

\bibitem{Witten4}
E.~Witten,
``The Verlinde algebra and the cohomology of the Grassmannian'',
arXiv:hep-th/9312104.

\bibitem{Gates1}
S.~J.~Gates, C.~M.~Hull and M.~Ro\v cek,
``Twisted multiplets and new supersymmetric nonlinear sigma models,''
Nucl.\ Phys.\ B {\bf 248} (1984) 157.

\bibitem{Spence1}
C.~M.~Hull, G.~Papadopoulos and B.~J.~Spence,
``Gauge symmetries for $(p,q)$ supersymmetric sigma models'',
Nucl.\ Phys.\ B {\bf 363} (1991) 593.

\bibitem{Hitchin1}
N.~Hitchin,
``Generalized Calabi-Yau manifolds'',
Q. J. Math.  {\bf 54}  (2003), no. 3, 281
[arXiv:math.DG/0209099].

\bibitem{Gualtieri}
 M.~Gualtieri,
``Generalized complex geometry'',
Oxford University Ph. D. Thesis, United Kingdom (2003),
arXiv:math.DG/0401221.

\bibitem{Lindstrom5}
U.~Lindstrom, M.~Ro\v cek, R.~von Unge and M.~Zabzine,
``Generalized Kaehler manifolds and off-shell supersymmetry'',
Commun.\ Math.\ Phys.\  {\bf 269} (2007) 833
[arXiv:hep-th/0512164].

\bibitem{Zayas1}
W.~Merrell, L.~A.~P.~Zayas and D.~Vaman,
``Gauged (2,2) sigma models and generalized Kaehler geometry'',
arXiv:hep-th/0610116.

\bibitem{Kapustin3}
A.~Kapustin and A.~Tomasiello,
``The general (2,2) gauged sigma model with three-form flux'',
JHEP {\bf 0711} (2007) 053 [arXiv:hep-th/0610210].

\bibitem{Witten1}
E.~Witten,
``Topological sigma models'',
Commun.\ Math.\ Phys.\  {\bf 118} (1988) 411.

\bibitem{Witten2}
E.~Witten,
``Mirror manifolds and topological field theory'',
in ``Essays on mirror manifolds'', ed. S.~T. ~Yau, International
Press, Hong Kong, (1992) 120 
[arXiv:hep-th/9112056].

\bibitem{Kapustin1}
A.~Kapustin,
``Topological strings on noncommutative manifolds'',
IJGMMP {\bf 1} nos. 1 \& 2 (2004) 49
[arXiv:hep-th/0310057].

\bibitem{Kapustin2}
A.~Kapustin and Y.~Li,
``Topological sigma-models with H-flux and twisted generalized complex manifolds'',
arXiv:hep-th/0407249.

\bibitem{Zucchini4}
R.~Zucchini,
``The biHermitian topological sigma model'',
JHEP {\bf 0612} (2006) 039
[arXiv:hep-th/0608145].

\bibitem{Zucchini5}
R.~Zucchini,
``BiHermitian supersymmetric quantum mechanics'',
Class.\ Quant.\ Grav.\  {\bf 24} (2007) 2073
[arXiv:hep-th/0611308].

\bibitem{Chuang}
W.~y.~Chuang,
``Topological twisted sigma model with H-flux revisited''
arXiv:hep-th/0608119.

\bibitem{Baptista1}
J.~M.~Baptista,
``Vortex equations in abelian gauged sigma-models'',
Commun.\ Math.\ Phys.\  {\bf 261} (2006) 161
[arXiv:math/0411517].

\bibitem{Baptista2}
J.~M.~Baptista,
``A topological gauged sigma-model'',
Adv.\ Theor.\ Math.\ Phys.\  {\bf 9} (2005) 1007
[arXiv:hep-th/0502152].

\bibitem{Baptista3}
J.~M.~Baptista,
``Twisting gauged non-linear sigma-models'',
arXiv:0707.2786 [hep-th].

\bibitem{Lindstrom2}
U.~Lindstrom, R.~Minasian, A.~Tomasiello and M.~Zabzine,
``Generalized complex manifolds and supersymmetry'',
Commun.\ Math.\ Phys.\  {\bf 257} (2005) 235
[arXiv:hep-th/0405085].

\bibitem{Lindstrom3}
U.~Lindstrom,
``Generalized complex geometry and supersymmetric non-linear sigma models'',
arXiv:hep-th/0409250.

\bibitem{Zucchini1}
R.~Zucchini,
``A sigma model field theoretic realization of Hitchin's generalized complex
geometry'', JHEP {\bf 0411} (2004) 045
[arXiv:hep-th/0409181].

\bibitem{Zucchini2}
R.~Zucchini,
``Generalized complex geometry, generalized branes and the Hitchin sigma model'',
JHEP {\bf 0503} (2005) 022
[arXiv:hep-th/0501062].

\bibitem{Pestun1}
V.~Pestun,
``Topological strings in generalized complex space'',
arXiv:hep-th/0603145.

\bibitem{Guttenberg}
S.~Guttenberg,
``Brackets, sigma models and integrability of generalized complex structures'',
JHEP {\bf 0706} (2007) 004
[arXiv:hep-th/0609015].

\bibitem{BV1}
I.~A.~Batalin and G.~A.~Vilkovisky,
``Gauge algebra and quantization'',
Phys.\ Lett.\ B {\bf 102} (1981) 27.

\bibitem{BV2}
I.~A.~Batalin and G.~A.~Vilkovisky,
``Quantization of gauge theories with linearly dependent generators'',
Phys.\ Rev.\ D {\bf 28} (1983) 2567
(Erratum-ibid.\ D {\bf 30} (1984) 508).

\bibitem{AKSZ}
M.~Alexandrov, M.~Kontsevich, A.~Schwartz and O.~Zaboronsky,
``The Geometry of the master equation and topological quantum field theory'',
Int.\ J.\ Mod.\ Phys.\ A {\bf 12} (1997) 1405
[arXiv:hep-th/9502010].

\bibitem{Zucchini3}
R.~Zucchini,
``A topological sigma model of biKaehler geometry'',
JHEP {\bf 0601} (2006) 041
[arXiv:hep-th/0511144].

\bibitem{Zucchini6}
R.~Zucchini,
``The Hitchin Model, Poisson-quasi-Nijenhuis Geometry and Symmetry
Reduction'', JHEP {\bf 0710} (2007) 075
[arXiv:0706.1289 [hep-th]].

\bibitem{Ikeda2}
N.~Ikeda,
``Two-dimensional gravity and nonlinear gauge theory'',
Annals Phys.\  {\bf 235} (1994) 435, 
[arXiv:hep-th/9312059].

\bibitem{Strobl}
P.~Schaller and T.~Strobl,
``Poisson structure induced (topological) field theories'',
Mod.\ Phys.\ Lett.\ A {\bf 9} (1994) 3129 
[arXiv:hep-th/9405110].

\bibitem{Cattaneo1}
A.~S.~Cattaneo and G.~Felder,
``A path integral approach to the Kontsevich quantization formula'',
Commun.\ Math.\ Phys.\  {\bf 212} (2000) 591 
[arXiv:math.qa/9902090].

\bibitem{Cattaneo2}
A.~S.~Cattaneo and G.~Felder,
``On the AKSZ formulation of the Poisson sigma model'',
Lett.\ Math.\ Phys.\  {\bf 56} (2001) 163
[arXiv:math.qa/0102108].

\bibitem{Witten5}
E.~Witten,
``On quantum gauge theories in two-dimensions'',
Commun.\ Math.\ Phys.\  {\bf 141} (1991) 153.

\bibitem{Witten6}
E.~Witten,
``Two-dimensional gauge theories revisited'',
J.\ Geom.\ Phys.\  {\bf 9} (1992) 303
[arXiv:hep-th/9204083].

\bibitem{Cartan1}
H.~Cartan,
``Notions d'alg\`ebre diff\'erentielle; applications aux groupes
de Lie et aux vari\'et\'es o\`u op\`ere un group de Lie'', 
Colloque de Topologie, C.B.R.M. Bruxelles (1950) 15.

\bibitem{Cartan2}
H.~Cartan,
``La transgression dans un group e de Lie et dans un espace fibr\`e principal'', 
Colloque de Topologie, C.B.R.M. Bruxelles (1950) 57.

\bibitem{Ratiu1}
J.~E.~Marsden and T.~S.~Ratiu,  
``Reduction of Poisson  manifolds'', 
Lett.\ in Math.\ Phys.\ {\bf 11} (1986) 161.

\bibitem{Vaisman}
I.~Vaisman, 
``Lectures on the Geometry of Poisson Manifolds'', 
Progress in Mathematics, vol. {\bf 118}, 
Birkhauser Verlag, Basel, Boston, Berlin (1994).

\bibitem{Ginzburg}
V.~Ginzburg,
``Equivariant Poisson cohomology and a spectral sequence associated with a
moment map'',  Int.\ J.\ Math.\ {\bf 10} (1999) 977
[math.DG/ 9611102]. 

\bibitem{Kirwan}
F.~C.~ Kirwan,
``Cohomology of Quotients in Symplectic and Algebraic Geometry'', 
Princeton University Press, (1984).

\bibitem{Bonechi}
F.~Bonechi and M.~Zabzine,
``Poisson sigma model on the sphere'',
arXiv: 0706.3164 [hep-th].

\bibitem{C-G-S} 
K.~Cieliebak, A.~R.~Gaio and D.~A.~Salamon,
``$J$-holomorphic curves, moment maps, and invariants of Hamiltonian
group actions'', 
Internat. Math. Res. Notices {\bf 16} (2000) 831 
[math.SG/9909122].

\bibitem{C-G-M-S} 
K.~Cieliebak, R.~A.~Gaio, I.~Mundet i Riera and D.~A.~Salamon,    
``The symplectic vortex equations and invariants of Hamiltonian group actions'', 
J. Symplectic Geom.  {\bf 1}  (2002) 543
[math.SG/0111176].

\bibitem{G-S} 
R.~Gaio and D.~A.~Salamon,
``Gromov-Witten invariants of symplectic quotients and adiabatic limits'',
J. Symplectic Geom.  {\bf 3}  (2005) 55
[math.SG/0106157]. 

\bibitem{MiR1} 
I.~Mundet i Riera,
Hamiltonian Gromov-Witten  invariants'',
Topology {\bf 42} (2003) 525  
[math.SG/0002121].

\bibitem{MiR2} 
I.~Mundet i Riera and G.~Tian,  
``Compactification of the moduli space of twisted holomorphic maps'', 
math.SG/0404407.  

\bibitem{MiR3}
I.~Mundet i Riera,
``Yang-Mills-Higgs theory for symplectic fibrations'',
U.A.M. Ph.D. Thesis, Spain (1999),
math.SG/9912150.

\bibitem{Calvo1}
I.~Calvo, F.~Falceto and D.~Garcia-Alvarez,
``Topological Poisson sigma models on Poisson-Lie groups'',
JHEP {\bf 0310} (2003) 033
[arXiv:hep-th/0307178].

\bibitem{Calvo2}
I.~Calvo and F.~Falceto,
``Dual branes in topological sigma models over Lie groups: BF-theory and
non-factorizable Lie bialgebras'',
JHEP {\bf 0604}, 058 (2006)
[arXiv:hep-th/0511212].

\bibitem{Bonechi2}
F.~Bonechi and M.~Zabzine,
``Poisson sigma model over group manifolds,''
J.\ Geom.\ Phys.\  {\bf 54} (2005) 173
[arXiv:hep-th/0311213].

\end{thebibliography}
\end{document}